\documentclass{aa}  

\usepackage{hyperref}
\usepackage{graphicx}
\usepackage{txfonts}
\usepackage{array}
\usepackage{placeins}
\begin{document} 
\setlength{\extrarowheight}{4pt}

   \title{The MeerKAT Fornax Survey} 
   \subtitle{IV. A close look at the cluster physics through the densest rotation measure grid}
   \titlerunning{A close look at the cluster physics through the densest rotation measure grid}

   \author{F. Loi\inst{1},
          P. Serra\inst{1},
          M. Murgia\inst{1},
          F. Govoni\inst{1},
          V. Vacca\inst{1}, 
          F. Maccagni\inst{1},
          D. Kleiner\inst{2,1} \and
          P. Kamphuis\inst{3}.
          }
    \authorrunning{F. Loi et al.}
   \institute{INAF - Osservatorio Astronomico di Cagliari,
              via della scienza 5, Selargius, Italy\\
              \email{francesca.loi@inaf.it}
        \and
            Netherlands Institute for Radio Astronomy (ASTRON), Oude Hoogeveensedijk 4, 7991 PD Dwingeloo, The Netherlands\\
        \and
            Ruhr University Bochum, Faculty of Physics and Astronomy, Astronomical Institute (AIRUB), 44780 Bochum, Germany
             }

   \date{Received September 15, 2023; accepted December 11, 2024}

  \abstract{
   Using the Square Kilometre Array (SKA) mid precursor MeerKAT, we acquired broadband spectro-polarimetric data in the context of the MeerKAT Fornax Survey to study the Fornax cluster's magnetic fields in detail by building the densest rotation measure (RM) grid to date. Here, we present the survey, the analysis, and a discussion of the RM grid properties.
   We analyzed a circular region centered on the Fornax cluster center with a radius of $\sim1.4^\circ$; that is, $\rm\sim 0.73 R_{vir}$. The mosaics have a resolution of 13$\arcsec$ and cover the frequencies between 900\,MHz and 1.4\,GHz, reaching an average noise of 16$\muup$Jy beam$^{-1}$ in total intensity and 3$\muup$Jy beam$^{-1}$ in the Q and U Stokes images. 
   With these data, we detected 508 polarized sources over an area of $\sim$6.35 deg$^2$ corresponding to a density of $\sim$80 polarized sources/deg$^2$. This is the densest RM grid ever built. Of the polarized sources, five are cluster sources. Excluding the cluster sources, we built the Euclidean-normalized differential source counts in polarization and we went a factor of ten deeper than previous surveys. We tentatively detect for the first time an increment in the differential source counts at low polarized flux densities; that is,  $\sim$9\,$\muup$Jy at 1.4\,GHz. The average degree of polarization of about 3--4\% suggests that the sub$-\muup$Jansky population is not dominated by star-forming galaxies, typically showing a degree of polarization lower than 1\%. The majority of the polarized sources are Faraday simple; in other words, their polarization plane rotates linearly with the wavelength squared. The RM shows the typical decrement going from the center to the outskirts of the Fornax cluster. However, interesting features are observed both in the RM grid and in the RM radial profiles across different directions. A combination of the cluster physics and large-scale structure filaments surrounding the Fornax cluster could explain the RM characteristics.}

   \keywords{magnetic fields -- polarization -- galaxies: clusters: individual: Fornax cluster -- surveys}

   \maketitle
%
%-------------------------------------------------------------------
\section{Introduction}
Among the most mysterious and elusive components of the Universe are the large-scale magnetic fields embedded in the intra-cluster medium of galaxy clusters and in the filaments of the cosmic web. 
A detailed reconstruction of such fields is crucial to probe the magneto-genesis mechanisms still under debate, to better understand the structure formation, and to build a complete picture of the physics and evolution of galaxies and galaxy clusters. 
In addition to the large-scale diffuse synchrotron emission associated with magnetic fields and relativistic particles in galaxy clusters \citep[see the review by][]{vanweeren2019}, the study of the Faraday effect on background polarized sources is a powerful tool to constrain their properties. The Faraday effect consists of the rotation of the polarization angle, $\Delta\Psi$, associated with a linearly polarized signal as a function of the wavelength squared $\lambda^2$:
\begin{equation}
\rm    \Delta\Psi = RM \cdot \lambda^2.
\end{equation}
The properties of the magnetic field are linked to the rotation measure (RM) as follows:
\begin{equation}
\rm    RM = \int_0^L B_{||} n_e dl,
\end{equation}
where n$_e$ is the thermal plasma density, B$_{||}$ is the line-of-sight parallel component of the magnetic field, and the integral is performed along the distance, L, between the observer and the source. 

At low frequencies ($\sim$100\,MHz), the evaluation of the RM helps characterize the strength of magnetic fields embedded in the filaments of the cosmic web \citep{osullivan2019A&A...622A..16O,osullivan2020MNRAS.495.2607O,Carretti2022,Carretti2023MNRAS}. With this technique, at mid-frequencies ($\sim$1\,GHz) it has been possible to measure the intra-cluster magnetic fields of several galaxy clusters \citep[e.g.][]{murgia2004, govoni2006, guidetti2008, laing2008, bonafede2010, guidetti2010, vacca2012, govoni2017, stuardi2021}. 
These studies show intra-cluster magnetic fields with a turbulent structure on a range of scales, from a few to hundreds of kiloparsecs. Their typical strength ranges from a few to a few tens of $\muup$Gauss. It is worth mentioning that the latter measurements are based on the RM of one or two sources up to a maximum of ten sources per cluster. Indeed, the sensitivity and resolution of the radio telescopes permit the detection of a limited number of polarized sources per square degree in a limited number of galaxy clusters. As a result, we lack a detailed reconstruction of the magnetic fields in clusters, which can help us to understand for example how intra-cluster magnetic fields evolve in different environments, what the turbulence scale is in clusters, and the impact of intra-cluster magnetic fields on the physics of cluster galaxies.\\
The Square Kilometre Array (SKA) would be the most suitable radio telescope to explore the injection, the evolution, and the impact of large-scale magnetic fields. Its precursors and pathfinders are already carrying out observations that are contributing to revolutionizing our knowledge of the Universe.  
Peculiar new sources and new features of known sources have been discovered, and these observations are challenging the physics of the objects populating the radio sky \citep[see e.g.,][]{Govoni2019,Botteon2020b,Botteon2020c,Ramatsoku2020A&A...636L...1R,rudnick2021,Riseley2022b,Vacca2022MNRAS.514.4969V,loi2023A&A...672A..28L}. 
A big step in this revolution is also expected in the cosmic magnetism research topic. In this sense, many surveys have been planned and are already ongoing, allowing us to build the so-called RM grid of the whole sky \citep[see][]{Heald2020Galax...8...53H}.
At mid-frequency, the POlarisation Sky Survey of the Universe's Magnetism \citep[POSSUM;][]{gaensler2010} with the Australian SKA Pathfinder \citep[ASKAP;][]{mcconnell2016,hotan} is ongoing. Early science publications show that this survey is detecting $\sim$25\,polarized sources per deg$^2$ from at a central frequency of 887 MHz using a 280\,MHz frequency band  \cite{anderson2021}. More recent results show a density of up to $\approx$37\,polarized sources per deg$^2$ when combining observations between 800--1087 MHz and 1316--1439\,MHz \citep{Vanderwoude2024AJ....167..226V}.
The Rapid ASKAP Continuum Survey recently released the Spectra and Polarisation In Cutouts of Extragalactic Sources \citep[SPICE--RACS]{Thomson2023PASA...40...40T}, which provides an area density of 4$\pm$2 polarized sources per deg$^2$ at 25$\arcsec$ between 744 and 1032\,MHz.
The MeerKAT International GigaHertz Tiered Extragalactic Exploration (MIGHTEE) recently released a catalog of 324 polarized sources between 880 and 1690 MHz in the COSMOS and XMM–LSS fields \citep{taylor2024MNRAS.528.2511T}.
In the future, the all-sky survey proposed to be performed with the SKA1--mid between 950 and 1760\,MHz is expected to detect 60 to 90 polarized sources per deg$^2$ at 2$\arcsec$ of resolution with a sensitivity of 4\,$\muup$Jy beam$^{-1}$ \citep{Heald2020Galax...8...53H}.\\
In this work, we present the broadband spectro-polarimetric data acquired within the MeerKAT Fornax Survey \citep[MFS]{serra2023}. It covers the Fornax galaxy cluster and the infalling Fornax A group. One of the goals of acquiring broadband spectro-polarimetric data is to map the Fornax cluster's magnetic field through the realization of the RM grid. This is the main focus of this work. 
Studying the intra-cluster magnetic field of Fornax is an important step in the investigation of large-scale magnetic fields in clusters for two main reasons: first, the cluster is a low-mass cluster, having a virial mass of M$_{\rm vir}$ = 5 $\times$ 10$^{13}$ M$_{\odot}$ \citep{Drinkwater2001ApJ...548L.139D} and a core radius of $\sim$173\,kpc \citep{chen2007A&A...466..805C}, and therefore it is representative of the majority of clusters in the Universe; second, it is nearby, at a distance of $\sim$20\,Mpc \citep{Blakeslee2001MNRAS.327.1004B,Jensen2001ApJ...550..503J,Tonry2001ApJ...546..681T}, meaning that we can study its magnetic field with unprecedented spatial resolution.\\
In a recent work, \citet{anderson2021} performed a RM grid of the Fornax cluster with the ASKAP telescope during the Early Science phase of POSSUM. These observations cover an area of $\sim$34\,deg$^2$ at 30\,$\mu$Jy beam$^{-1}$ of sensitivity with a resolution of 10$\arcsec\times$14$\arcsec$. With an average density of 25 polarized sources per deg$^2$, these authors show the presence of a background scatter of about 17\,rad m$^{-2}$ in the RM within 360 kpc from the cluster center; that is, 2--4 times larger than the X-ray emitting plasma radius. In addition, the enhancement in RM is not symmetric but rather points along the merger axis between the Fornax cluster center and the infalling Fornax A group. The authors suggest that the merger could explain the peculiarities observed in the RM grid.\\
With the MFS, we have detected the polarized signal in the Fornax cluster field with even higher sensitivity.
Indeed, with the MFS broadband spectro-polarimetric data, we reconstructed the densest RM grid ever built so far: $\sim$80 polarized sources per deg$^2$. Thanks to this, we can better characterize the cluster physics.\\
In Sect. 2, we describe the survey, data reduction, and imaging. In Sect. 3, we explain how we derive the polarized and RM images from Q and U Stokes cubes, and we discuss the cluster polarized sources and the differential polarized source counts. In Sect. 4, we discuss the RM properties. Sect. 5 reports a discussion of the results and the conclusion are in Sect. 6.
At the distance of the Fornax cluster, 1$\arcsec$ corresponds to 0.1 kpc.

%--------------------------------------------------------------------
\section{The survey}
The MFS has already been described in detail in \cite{serra2023}. Here, we consider the 4K broadband full-Stokes data acquired simultaneously with the 32K-zoom band data presented and analyzed by \cite{serra2023,kleiner2023A&A...675A.108K,Loni2023}. The data flagging and calibration performed here follow the same steps described in \cite{serra2023} for the 32K zoom band data. Therefore, we shall briefly describe the broadband data used in this work and how we calibrated them in total intensity, while we shall carefully illustrate the polarization calibration and imaging.

\subsection{Observations}
The 4K broadband full-Stokes data were acquired between 856\,MHz and 1712\,MHz with a channel width of 208.984\,kHz. As is reported in Table 1 of \cite{serra2023}, we observed a primary calibrator (either 1934-638 or 0408-658), which we also used to calibrate the on-axis leakage, a secondary calibrator (J0440-4333), and a polarized calibrator (3C138). Here, we present the data reduction and imaging of 50 pointings covering the cluster central region. More details about the survey design are reported in \citep{serra2023}.\\

\subsection{Data reduction} 
We performed the data reduction with the \texttt{CARACal} pipeline \citep{josh2020}. First, we split the measurement set to obtain a separate set with the calibrators. After first flagging the autocorrelations, shadowed antennas, and the frequency ranges
1408--1423\,MHz and 1419.5--1421.3\,MHz (where the 21 cm-wavelength neutral hydrogen emission or absorption from, respectively, the Fornax cluster and the Milky Way can corrupt the band-pass calibration), we adopted automatic flagging procedures to excise the remaining radio frequency interferences (RFI). In particular, we made use of the \texttt{AOFlagger} \citep{offringa2012} tool based on the Stokes QUV visibilities. Then, we calibrated the cross-hand correlation (HH, VV) excluding baselines shorter than 100 m. The calibration tables were derived twice. Indeed, after a first round of solutions, we applied the tables and refined the RFI excision with the \texttt{CASA} task {\tt tfcrop}. The portion of flagged data is around 60\%. We excluded the data below 900\,MHz and above 1.65\,GHz to avoid band-pass roll-offs. \\
To calibrate the data in polarization, we made use of the \texttt{CARACal} pipeline, where standard \texttt{CASA} tasks have been implemented in the \texttt{polcal} worker.
The pipeline includes three strategies to calibrate the polarization, depending on the calibrators available during the observations (see the Appendix \ref{app:polcal} for more details). We could observe both an unpolarized and a known polarized calibrator during each observation. Therefore, we solved for the cross-hand delay (KCROSS) and for the phase (Xf) with 3C138, then for the off-axis leakage (Df) with either J1939-6342 or J0408-6545.
The spectro-polarimetric properties of the polarized calibrator were taken from the NRAO website\footnote{https://science.nrao.edu/facilities/vla/docs/manuals/obsguide/modes/pol, Table 7.2.7} and from \cite{perley2017}.\\
Despite the strategy used to calibrate the data, we noted the presence of a polarization angle offset of $\sim$7\,deg at 1\,GHz both in the MFS data and in the commissioning data of the MFS. This offset appears to be stable across the frequency band. Private communications with astronomers calibrating MeerKAT data with tools different from \texttt{CARACal} or CASA confirmed the presence of such offset. We stress that this is not relevant, since the analysis presented in this work is not based on the absolute polarized angle value.

\subsection{Target flagging, imaging, and final mosaics}
We split the target from the original measurement set with \texttt{CARACal} and we simultaneously applied the calibration tables. We flagged the frequency ranges mentioned in the previous section, the autocorrelation, and the antenna shadows, and we used \texttt{tricolour}\footnote{\url{https://github.com/ratt-ru/tricolour}} to excise the remaining RFIs.
The percentage of data flagged is around 50-60\%.

In total intensity, we imaged the data between 900\,MHz and 1.65\,GHz to avoid the band-pass roll-off using the \texttt{CARACal} tool. The imaging worker loops between the imaging of the data, performed with \texttt{WSclean} \citep{offringa2014,offringa2017}, and the self-calibration, with \texttt{Cubical} \citep{kenyon2018}.
We decided to image the data in a 4096 pixel grid using a pixel size of 2.5$\arcsec$. We enabled the "join-channels" of \texttt{WSclean} to improve the cleaning and image the data into five output channels. We used the Briggs weighting scheme with $robust$=0 and a 6$\arcsec$ $uv-$tapering. We tried to capture the spectral behavior across the bandwidth by fitting a fourth-order polynomial. \\
After a first imaging step by using the automatic mask made by \texttt{WSclean}, we used the \texttt{SoFiA} tool \citep{serra2015} to improve it. The noise was evaluated in local windows with a 100 pixels width. The data were imaged again using the \texttt{SoFiA} mask. The self-calibration was then performed with \texttt{cubical} using the “Fslope” scheme that calibrates the delay in chunks of 2 minutes each across the entire bandwidth. We performed the self-calibration and imaging loop three times. At each round, we made a mask with \texttt{SoFiA} considering 8, 5, and 4 rms thresholds to identify the sources.
\begin{figure*}
    \centering
    \includegraphics[width=0.9\textwidth]{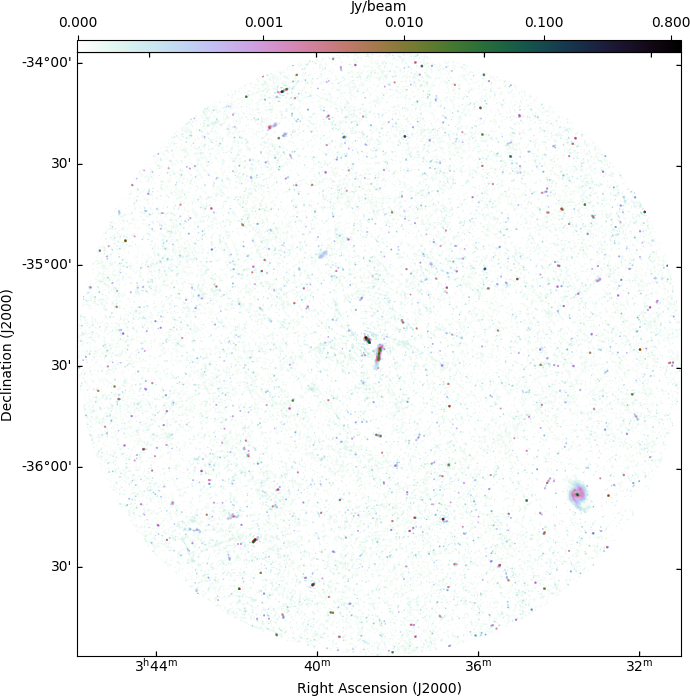}
    \caption{1.15\,GHz total intensity emission average between 900\,MHz and 1.4\,GHz of the central part of the Fornax cluster. The average noise in the image is $\sim$16$\muup$Jy beam$^{-1}$. The resolution is 13$\arcsec$. The color bar is on a logarithmic scale.}
    \label{fig:ti}
\end{figure*}

We imaged again the self-calibrated measurement set with the \texttt{polcal} worker. In particular, we created the Stokes Q and U cubes considering the frequency interval between 900\,MHz and 1.4\,GHz recommended by SARAO to minimize the off-axis leakage. The frequency resolution of the cubes is 5\,MHz. We enabled the “join-channels,” “squared-channel-joining,” and the “join-polarizations” to improve the cleaning.
We then convolved the frequency cubes to a common resolution of 13$\arcsec$ and computed the primary beam cubes assuming the model by \cite{mauch2020}. We made the Stokes I, Q, and U mosaic with \texttt{MosaicQueen}\footnote{https://github.com/caracal-pipeline/MosaicQueen} using a primary beam cutoff of 10\%.

Figure \ref{fig:ti} shows the mean total intensity image between 900\,MHz and 1.4\,GHz, with a resolution of 13$\arcsec$. The field of view is centered on the cluster center and its diameter is $\sim$2.8$^\circ$.
The average noise in the image, computed as the median absolute deviation of all the negative pixels, is $\sigma$=16\,$\muup$Jy beam$^{-1}$.

\section{Polarized intensity image}
In this section, we show the polarized intensity image obtained after the application of the RM synthesis technique \citep{burn1966,BrentjensdeBruyn} on the Stokes Q and U mosaics. In addition, we discuss the properties of the five cluster sources that show an associated polarized emission. Finally, we report the Euclidean-normalized polarized source counts detected in the field of view.

\subsection{Rotation measure synthesis technique and positive bias subtraction}
We applied the RM synthesis technique on the Stokes Q and U mosaic cubes to overcome the bandwidth depolarization. We used \texttt{RMtools} \citep{rmtools} and in particular the \texttt{RMsynth3d} routine to apply the technique between $-$200\,rad m$^{-2}$ and 200\,rad m$^{-2}$ using a step of 2\,rad m$^{-2}$. Considering the selected data, the corresponding FWHM in Faraday space, the largest RM scale and the maximum RM value detectable are, respectively:
\begin{eqnarray}
    \delta \phi &\approx& \frac{2\sqrt{3}}{\Delta \lambda^2} \approx 59\,\rm rad/m^2\\ \nonumber
    \rm max-scale &\approx& \frac{\pi}{\lambda_{\rm min}^2}  \approx 68\,\rm rad/m^2 \\ \nonumber
    |\phi_{\rm max}| &\approx& \frac{\sqrt{3}}{\delta\lambda^2} \approx 2332\,\rm rad/m^2 .
\end{eqnarray}
Following \cite{Rudnick2023MNRAS.522.1464R}, we also computed the Faraday width, W$_{max}$, at which the power in the Faraday spectrum drops by a factor of 2 for single Faraday components:
\begin{equation}
    \rm W_{max} = 0.67 \left ( \frac{1}{\lambda_{min}^2} + \frac{1}{\lambda_{max}^2} \right ) \approx 26\,rad/m^2 .
\end{equation}

We assumed that the sources are Faraday simple (i.e., consist of a single component at a given Faraday depth) so that the polarized intensity image corresponds to the maximum polarized signal in the Faraday dispersion function (FDF). This aspect will be better discussed in Sect. 4.1.

\begin{figure*}
\centering
\includegraphics[width=0.92\textwidth]{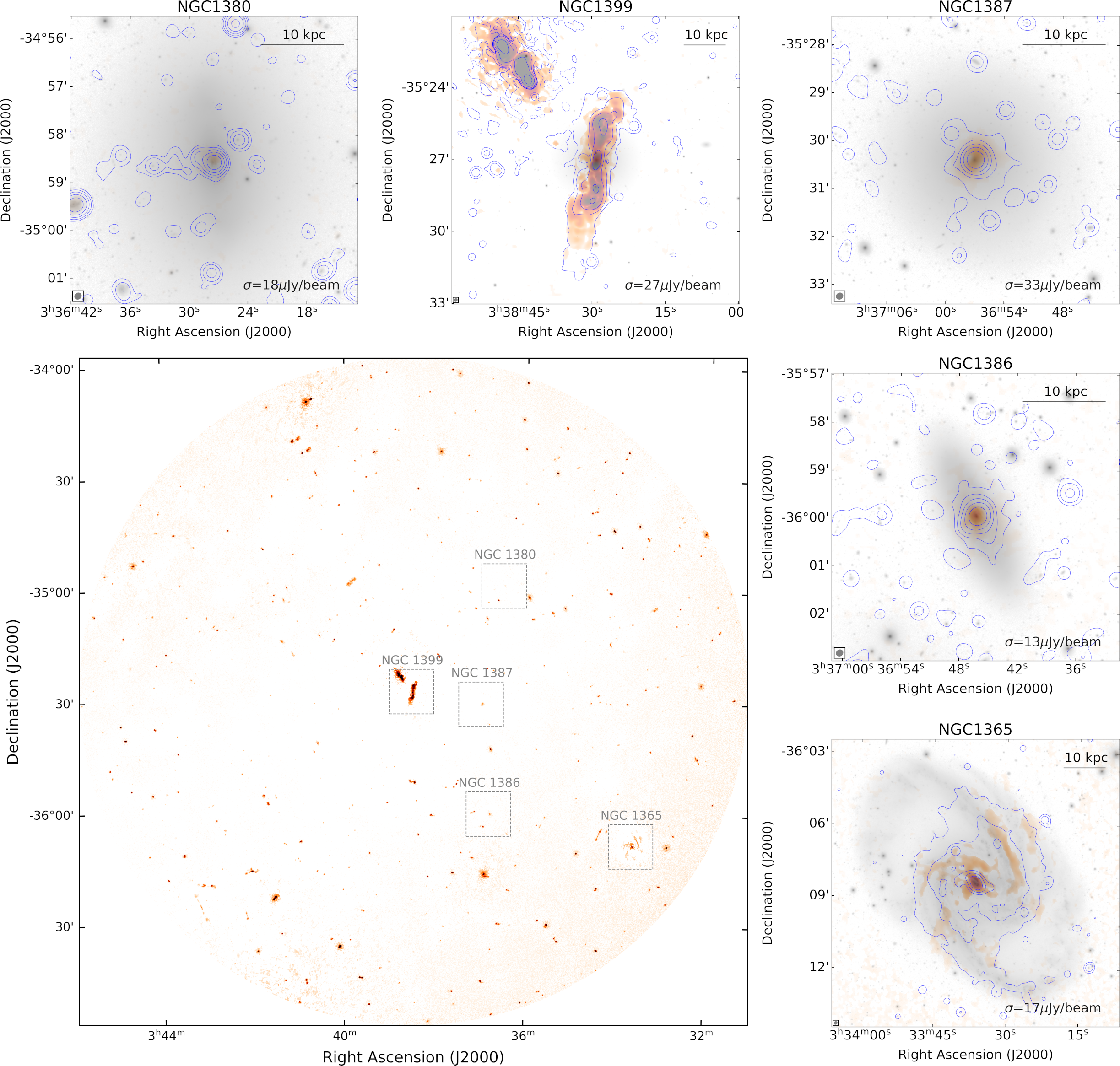}
    \caption{Polarized intensity after the application of the RM synthesis and the removal of the positive bias obtained between 900\,MHz and 1.4\,GHz. The resolution is 13$\arcsec$. The zoomed images around the bottom left image are centered on the five polarized cluster sources. For each of them, we show the optical image from the Fornax Deep Survey in gray scale and the total intensity contours starting from 3$\sigma$, whose value is reported in the bottom right corner of the image, and increasing on a logarithmic scale. }
    \label{fig:pol}
\end{figure*}

The polarized intensity image suffers from the positive bias due to the propagation of the Q and U noise. We removed it following these steps:
\begin{enumerate}
    \item we ran \texttt{SoFiA} on the (bandwidth-averaged, considering RM=0) Q and U images to mask the most obvious sources in the field;
    \item using the previous mask as input, we ran \texttt{SoFiA} again to derive an image of the noise, both for Q and U Stokes parameters;
    \item we created a QU noise image by averaging the Q and U noise images; the result is an average $\rm\sigma_{QU}=(\sigma_{Q}+\sigma_{U})/2=2.6\,\muup$Jy beam$^{-1}$ with a minimum of 1.4\,$\muup$Jy beam$^{-1}$ and a maximum of 19\,$\muup$Jy beam$^{-1}$.
    \item following Eq. 5 of \citep{george2012}, we derived the debiased polarized intensity image as $\rm \sqrt{P^2-2.3\sigma_{QU}^2}$, where P is the polarized surface brightness.
\end{enumerate}
Fig. \ref{fig:pol} shows (bottom left) the debiased polarized intensity image obtained between 900\,MHz and 1.4\,GHz in a circular region 1.42$^\circ$ in radius around the cluster center.\\
The majority of the sources in the field are point-like. However, we can see many spectacular extended objects both within the cluster, such as the radio jets associated with the brightest cluster galaxy (BCG) NGC\,1399 or the spiral galaxy NGC\,1365, and in the field.\\
To derive a catalog of the polarized sources and the corresponding RM grid, we followed these steps:
\begin{enumerate}
    \item we de-rotated the Q and U images, assuming as RM the Faraday depth at the maximum FDF for each line of sight;
    \item we ran \texttt{SoFiA} on the de-rotated Q and U images to create two masks of the Q and U detections. We set a threshold of 5$\sigma$, where $\sigma$ was evaluated in running windows with a 50 pixel width;
    \item we joined the two masks to be sure to include sources with detectable Q and/or U signals;
    \item we ran \texttt{SoFiA} again with the previous image as an input mask to extract a mask and a catalog on the debiased polarized intensity image;
    \item we used the debiased polarized mask to isolate the RM measurements.
\end{enumerate}
To be sure that we were not contaminated by spurious detection, we ran the \emph{SoFiA} tool on the total intensity image to compare the output catalog with the polarized intensity one. We created a total intensity catalog, discarding the sources with a reliability of less than 99.2\%, an integrated SNR<7, and a size smaller than the beam size. The resulting catalog has a reliability equal to 99.99\%. This was computed as the ratio between the detections minus false detections over the detections. The number of false detections is the sum of 1 minus the source reliability computed on all the sources in the catalog. We compared this catalog with the polarized catalog to be sure that every polarized source has a total intensity counterpart. After this check and a visual inspection of the images, we discarded one source from the initial polarized intensity catalog. However, after a visual inspection we removed one more source that was likely due to the presence of an artifact.
As a result, we identify 508 polarized sources in the debiased polarized intensity image, which amounts to a source density of about 80 polarized sources per deg$^2$. This is the densest polarized grid ever built so far and it is comparable with the expectation for the Square Kilometre Array \citep[see][]{Heald2020Galax...8...53H}.

\subsection{Source identification and distribution}
We compared our detections with the spectroscopic catalog by \cite{maddox}. We found that five objects are cluster sources \citep[out of 264 confirmed members within 1.42$^{\circ}$ from the cluster center, see][]{venhola}; namely,  NGC\,1380, NGC\,1399, NGC\,1387, NGC\,1386, and NGC 1365. We show the zoomed images centered on these sources (in descending order of declination) in Fig. \ref{fig:pol}. In these images, we also show the optical Fornax Deep Survey image \citep{iodice,venhola} in gray scale, the polarized intensity in red color, and the total intensity contours in blue. 

The five cluster sources that show polarized intensity emission are all found west of the cluster center. Here, we briefly summarize their properties.

\subsubsection*{NGC\,1380}
NGC\,1380 is an early-type galaxy and an ancient infaller with only traces of star formation activity in the center \citep{iodice2019}. Its radio counterpart at 1.15\,GHz in total intensity shows a compact core and an extended emission toward the east. In polarization, it is a point-like source with a size coincident with the dust and CO disc \citep{sarzi2018,zabel2019}. The radio continuum emission could be associated with thermal emission. This is compatible with the observed polarization degree, which is less than 1\% \citep{Stil2009}. However, part of the radio emission could be associated with the central AGN \citep{Viaene2019}.

\subsubsection*{NGC\,1399}
NGC\,1399 is the BCG of the Fornax cluster \citep{iodice}. It is a massive elliptical galaxy showing two elongated jets \citep{Killeen1988ApJ...325..180K} in radio continuum and polarization with a total extension of about 50\,kpc.\\
On average, it shows a polarization degree of $\sim$4\%, reaching levels of $\sim$20\% in the southern part (between the lower and second contours shown in the zoomed inset in Fig. \ref{fig:pol}). 

\subsubsection*{NGC\,1365}
NGC\,1365 is a spiral galaxy with a radio continuum counterpart that is extremely faint in polarization. Indeed, we can only detect polarized signals at the core of the galaxy and some patches (below 5$\sigma$) in the intra-arm regions. It has a size of about 50\,kpc in total intensity and 8\,kpc in polarization, considering a 3$\sigma$ and a 5$\sigma$ threshold, respectively. Its average polarization degree is below 1\% and increases far from the core of the galaxy up to $\sim$6\%. \cite{beck2002} found an average polarization degree of (1.6$\pm$0.4)\% at 1.36\,GHz and a RM at the center of $\approx-$20\,rad m$^{-2}$ (computed as a linear fit between two frequencies), while we have an average RM of about $-$15$\pm$1\,rad m$^{-2}$.

\subsubsection*{NGC\,1386}
NGC\,1386 is a Seyfert 2 early-type spiral galaxy \citep{rodriguez2017} and one of the recent infallers of the cluster \citep{iodice2019}.
It has a largest linear size of about 14\,kpc and 4.6\,kpc in total intensity and polarization, respectively. CO and H$\alpha$ disks are detected at the same scales as the polarized intensity \citep{zabel2019,iodice2019}.
Its average polarization degree is below 1\% and increases toward the south far from the core of the galaxy up to $\sim$5\%. The radio continuum emission could be due to both thermal and nonthermal components in the galaxy.

\subsubsection*{NGC\,1387}
NGC\,1387 is an early-type galaxy and an ancient infaller showing a circumnuclear star formation ring \citep{iodice2019}. Its radio counterpart covers a projected linear size of about 5\,kpc, which goes beyond the H$\alpha$ and dust ring \citep[<2\,kpc]{iodice2019}. It has an associated polarization degree of 4\% on average that increases from the galaxy center in the NW and SE directions ($\sim$55\,deg with respect to WE, perpendicular to the stellar rotation axis) up to $\sim$16\%. These could be radio jets.

\subsection{Differential polarized source counts}
\begin{figure}
\centering
    \includegraphics[width=0.4\textwidth]{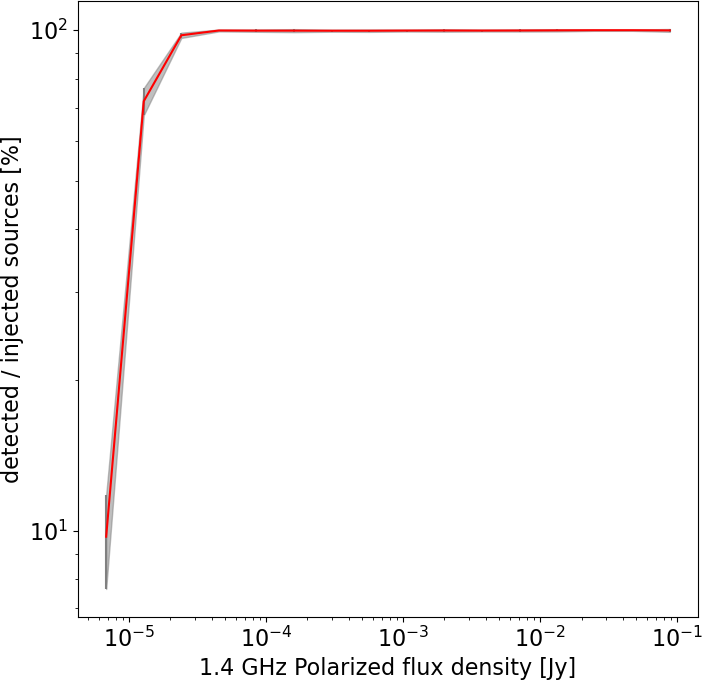}
    \caption{Completeness as a function of the polarized flux density computed with simulations, see the text for more details.}
    \label{fig:completeness}
\end{figure}

We evaluated the differential source counts in polarization in the debiased polarized intensity image, only in this part of the work we excluded the five cluster polarized sources. Our goal here is to point out the consistency of our measured polarized intensity with previous surveys and to show the sensitivity level achieved. \\
First of all, we need to compute the completeness factor at different polarized intensities. This correction is particularly important also because the QU noise is not uniform across the field of view. Therefore, going to low polarized intensities, the number of sources that can be detected becomes smaller and smaller. To account for this effect, we performed a simulation. Once we verified that the faintest sources are point-like sources, we simulated 100 point-like sources by randomly extracting the polarization angle and the polarized flux intensity within a given range of polarized flux density. After deriving the Q and U signals of the simulated sources, we randomly placed them in the de-rotated Q and U Stokes images, repeating the extraction if the distance from another (simulated or real) source was less than 40 pixel ($\sim$ 14\arcsec).
We performed the source detection following the same steps described in Sect. 3.1. For each bin of polarized flux density, we repeated the procedure 30 times. 
We evaluated the percentage of detected sources over the injected ones and corrected the source counts for this value.\\
Fig. \ref{fig:completeness} shows the completeness as a function of the polarized flux density rescaled at 1.4 GHz considering a spectral index of $\alpha$=0.8, assuming that the polarized flux density, P$\rm_{\nu}$, scales with the frequency, $\nu$, as P$\rm_{\nu}\propto\nu^{-\alpha}$. The dashed area shows the uncertainty that corresponds to the standard deviation of the 30 distributions per bin of polarized intensity. \\
\begin{table}
\caption{Euclidean-normalized differential polarized source counts.}
\label{table:dnds}
\centering
\begin{tabular}{c c c c c}
\hline\hline
P$\rm_{1.4\,GHz}$ & $\rm\Delta P_{1.4\,GHz}$ & N & N$_{\rm corr}$ & dN/dP P$^{2.5}$  \\  
mJy & mJy &  &  & Jy$^{1.5}$sr$^{-1}$ \\
\hline
   0.009 & 0.007--0.013 & 49 & 503 & 0.012$^{+0.003}_{-0.004}$\\
   0.018 & 0.013--0.024 & 122 & 169 & 0.010$^{+0.001}_{-0.001}$\\ 
   0.033 & 0.024--0.045 & 103 & 106 & 0.016$^{+0.002}_{-0.002}$\\ 
   0.062 & 0.045--0.085 & 58 & 58 & 0.023$^{+0.003}_{-0.003}$\\
   0.117 & 0.085--0.160 & 46 & 46 & 0.047$^{+0.008}_{-0.007}$\\   
   0.219 & 0.16--0.30 & 41 & 41 & 0.11$^{+0.019}_{-0.017}$\\
   0.412 & 0.30--0.57 & 35 & 35 & 0.24$^{+0.047}_{-0.039}$\\
   0.774 & 0.57--1.06 & 24 & 24 & 0.42$^{+0.10}_{-0.084}$\\
   1.46  & 1.06--2.00 & 10 & 10 & 0.45$^{+0.19}_{-0.14}$\\
   2.74  & 2.00--3.75 & 8 & 8 & 0.92$^{+0.46}_{-0.32}$\\
   5.15  & 3.75--7.06  & 3 & 3& 0.89$^{+0.87}_{-0.48}$\\
   9.68  & 7.06--13.3  & 0 & 0 & 0\\
   18.2  & 13.3--25.0 & 3 & 3& 5.9$^{+5.8}_{-3}$\\
   34.2  & 25.0--46.9 & 0 & 0 & 0\\
   64.3  & 46.9--88.2 & 1 & 1& 13$^{+30}_{-11}$\\
\hline
\end{tabular}
\end{table}

\begin{figure*}
\centering
    \includegraphics[width=0.8\textwidth]{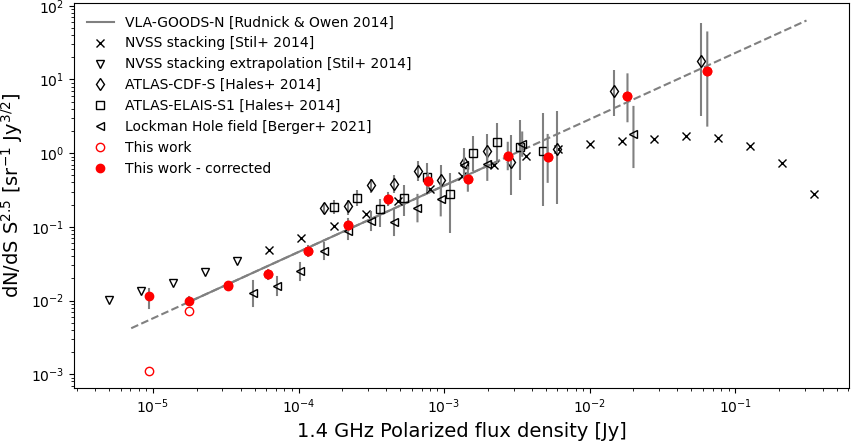}
    \caption{Euclidian-normalized differential source counts at 1.4 GHz in polarized intensity from our survey (red circles). We also show the data taken from \cite{rudnick2014} (the solid line represents the function fitted on the VLA-GOODS-N field data, the extrapolation at lower and higher polarized flux densities is shown as dashed line), \cite{hales2014} (diamonds and squares refer to the ATLAS-CDF-S and the ATLAS-ELAIS-S1 fields, respectively), \cite{berger2021} (left-facing triangles for the Lockman Hole field), \cite{stil2014ApJ...787...99S} (crosses for the stacking of the NVSS sources and downward-pointing triangles for the extrapolation at lower polarized flux densities).} 
    \label{fig:dnds}
\end{figure*}
The 1.4\,GHz differential polarized source counts are reported in Table \ref{table:dnds}. We multiplied the polarized differential source counts by the polarized flux density, P$\rm_{1.4\,GHz}^{2.5}$, to obtain normalized counts considering an Euclidean Universe. In a static Universe, the normalized source counts should follow an horizontal line as a function of the flux density, while in an expanding Universe with a constant comoving source density we should observe a monotonic decrease going at low flux densities. A flattening at low flux density would suggest the cosmological evolution of the radio sources.
In each column of Table \ref{table:dnds}, we indicate the central polarized flux density bin, P$\rm_{1.4\,GHz}$, and the interval $\Delta$P$\rm_{1.4\,GHz}$ at 1.4\,GHz in mJansky, the number, N, of polarized sources per bin, N corrected for the completeness, N$_{\rm corr}$, and the value of the Euclidean-normalized differential polarized source counts (considering N$_{\rm corr}$), respectively.
Also, in this case we considered $\alpha$=0.8 to rescale the polarized flux density. We calculated the Poissonian error on the number of sources detected in each beam following \citep{regener1951}. In the low polarized intensity regime probed for the first time in our survey, an additional source of uncertainty is due to the one associated with the completeness. Therefore, we also added in quadrature its standard deviation to the uncertainties. These are reported in Table \ref{table:dnds} in column 5.
Fig. \ref{fig:dnds} shows the Euclidean-normalized differential polarized source counts. The empty and filled red dots are the observed and completeness corrected counts, respectively. At large polarized flux density, they overlap, since the completeness tends to 100 percent.
We include in the plot the results from several surveys in polarization: the ATLAS--CDF--S and the ATLAS--ELAIS--S1 \citep{hales2014}, and the Lockman Hole field \citep{berger2021} plotted as diamonds, squares, and left-facing triangles, respectively. 
The fields of view of these surveys are 3.626\,deg$^2$, 2.766\,deg$^2$, and 6.5\,deg$^2$, respectively.
Our field of view is 6.35\,deg$^2$ and, at variance with the previous ones, we do not see edge effects due to the primary beam since we are considering a subregion of the entire mosaic. 
We have not included the recent polarized source counts at 140\,MHz \citep{piras2024A&A...687A.267P} as the depolarization effects in dense environments at such low frequencies would make the comparison unfair.
The solid line represents the empirical function derived by \cite{rudnick2014} in the VLA--GOODS--N data (field of view of 0.275\,deg$^2$). According to these authors, at 10\arcsec of resolution the cumulative number of polarized sources per square degree scales as a function of the polarized flux density as:
\begin{equation}
    N(>p) = 48 \cdot (p/\rm 30\,\muup Jy)^{-0.6}.
\end{equation}
We considered the results obtained removing the three high-RM sources \citep[for more details see][]{rudnick2014}.
The dashed line is the extrapolation at lower and higher polarized flux densities of the latter function.
We also plot the result of the NVSS source stacking by \citep{stil2014ApJ...787...99S} as crosses and their extrapolation at lower polarized flux densities as downward-pointing triangles.\\
The distribution of our points is not homogeneous with polarized intensity. This is expected at high polarized intensity, since sources in this regime are rare in the Universe. Indeed, the number of polarized counts in the last five bins is seven. In any case, we find a good agreement with the previous surveys, especially with what is reported by \cite{berger2021} below 20\,mJy. At low polarized flux density, the differential polarized source counts decrease. We can see a good agreement with the prediction by \cite{rudnick2014}. However, in the lowest polarized flux density bin at 9\,$\muup$Jy we see a difference of 2.2 dex with respect to the empirical function by \cite{rudnick2014}. We compare such a difference with the uncertainty related to our data and we find that the deviation is about 1.5 times the uncertainty at 9\,$\muup$Jy. We also notice that our measurement is closer to the extrapolation by \cite{stil2014ApJ...787...99S}, while up to about 0.1\,mJy these authors report higher values compared to our work. It is worth mentioning that the empirical function derived at 1.6$\arcsec$ by \cite{rudnick2014} is compatible with the work by \cite{stil2014ApJ...787...99S}. At 62\,$\muup$Jy, our measurement is below the estimates reported by \cite{rudnick2014} (-2.2 dex), with a discrepancy of about 2.3 times the uncertainty. Nevertheless, our value at 62\,$\muup$Jy is compatible with the work by \cite{berger2021}. \\ 
We note that the average degree of polarization at low polarized flux densities is around 3-4\%. Such a value is more compatible with an emission triggered by AGN rather than associated with star-forming galaxies, since the latter sources show in average a degree of polarization of about 0.8\% \citep{Stil2009,taylor2014ASInC..13...99T,Bonaldi2019MNRAS.482....2B}.

%-----------------------------------------------------------------
\section{Rotation measure properties}
In this section, we show the RM image and analyze its properties. In particular, we demonstrate that the majority of the sources are Faraday simple; in other words, their polarized signal crosses an external Faraday screen and there is no hint at this level of internal rotation. Eventually, we show the radial profiles of the RM, which evidence the presence of a strong asymmetry. 
\begin{figure*}
    \centering
    \includegraphics[width=0.8\textwidth]{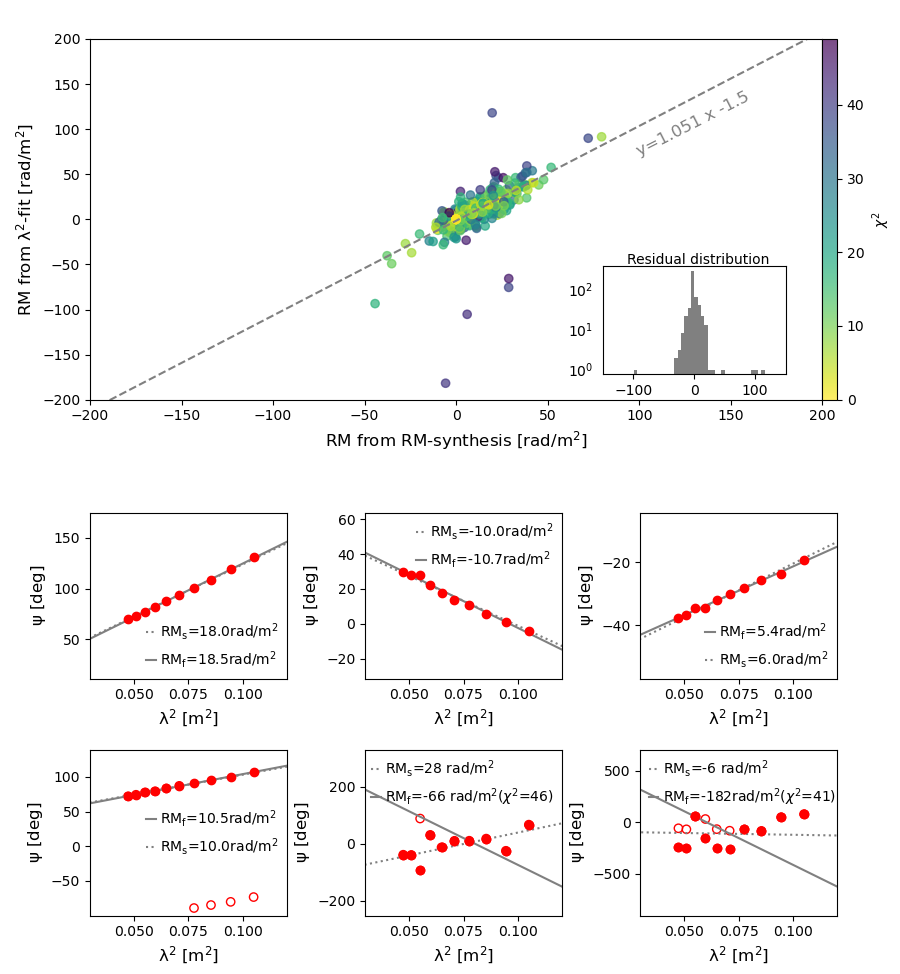}
    \caption{Faraday complexity analysis. Top: Comparison between the RM derived with the RM synthesis (x axis) and the $\lambda^2-$fit techniques applied to the entire mosaic. Each point is the RM at the polarized surface brightness peak of each source. The points are color-coded according to the reduced $\chi^2$ of the $\lambda^2-$fit. The bottom right inset shows a histogram of the residuals. Middle and bottom: The red points show the polarized angle as a function of $\lambda^2$. The empty and filled points are, respectively, the observed and the n$\pi-$ambiguity-corrected points. In each plot, the results of the RM synthesis and of the $\lambda^2-$fitting techniques are reported as dotted and solid lines, respectively. The RM values are also reported in each plot with the value of the reduced $\chi^2$ when larger than 1.}
\label{fig:rmlinearity}
\end{figure*}
\subsection{Faraday complexity}
As we already stated in the introduction, the RM is defined as the constant of proportionality between the rotation of the polarization plane of a background source and the wavelength squared. 
In the particular case of an internal RM contribution -- an emitting and rotating source -- the $\rm\lambda^2-$law breaks and we can observe a complex FDF; in other words, multiple components in the FDF. 
Faraday simple sources, which by definition show a single peak in the FDF, can sample a Faraday depth that corresponds to the RM associated with the intervening medium (ICM, IGM, Galactic RM, cosmic filaments RM, etc.). At gigahertz frequencies and in the direction of the Fornax cluster, we expect to detect a major contribution from the ICM and the Galactic foreground. 
It is therefore important to check if the majority of the polarized sources are Faraday simple. If this is true once we subtract the Galactic RM from the observed RM (i.e., the Faraday depth at the peak of the FDF), we are confident that we can obtain a reliable estimate of the intra-cluster magnetic field properties.\\

To understand if the majority of the sources are Faraday simple, we compare the observed RM with the result of a $\lambda^2-$fitting procedure on the observed polarized angle. To perform the fit, we used the \texttt{FARADAY} tool \citep{murgia2004}.\\
We binned the Q and U frequency cubes in 10 channels of 50\,MHz each. Then, we evaluated the polarization angle at each channel and fit the data pixel by pixel with a least square fitting procedure. The algorithm can mitigate the n$\pi-$ambiguity with a similar method used in the \texttt{PACERMAN} tool \citep{dolag2005MNRAS.358..726D}: first we determined the RM for the pixels that have the highest signal-to-noise and then we used that RM as an initial guess for nearby pixels.\\
The comparison between the RM evaluated with the RM synthesis and the $\lambda^2-$fit is displayed in Fig. \ref{fig:rmlinearity} on the top. The scatter plot is color-coded according to the reduced $\chi^2$ evaluated in the $\lambda^2-$fitting procedure. Each point in the plot represents a polarized source: for each point-like source, we selected the line of sight with the maximum polarized surface brightness across the source itself. For the extended sources, we considered a number for the line of sight corresponding to the number of beams covered by the source area.\\
The majority of the sources show the same RM regardless of the technique used to determine it. This can be further appreciated in the inset plot that shows the histogram of the differences between the RM derived with the RM synthesis and the $\lambda^2-$fitting techniques. The y axis is on a logarithmic scale. We note that the few outliers show a high reduced $\chi^2$.
We evaluated the linearity of the relation between the RM derived with the RM synthesis and the $\lambda^2$ fitting techniques using a least square fitting that suggests a slope of m=1.051$\pm$0.002 and a y intercept of b=-1.5$\pm$0.5. This means that the majority of the polarized sources are Faraday simple. Some examples are reported in the middle and bottom panels of the figure. The three middle plots and the bottom left plot show a perfect match between the RM derived with the RM synthesis and the $\lambda^2$ fitting techniques, represented by the dashed and solid lines, respectively. The binned data are shown as red points. The empty points here are the observed ones, while the filled points are corrected for the n-$\pi$ ambiguity.
The last two bottom plots of Fig. \ref{fig:rmlinearity} show two outliers of the top plot, with $\chi^2$ larger than 40. These plots show a case where the RM determined with the two methods is significantly different. 
This could be due to the difficulty in overcoming the n$\pi-$ambiguity in the case of the $\lambda^2-$fit. It is indeed well known that the RM synthesis technique can easily accommodate this issue \citep{BrentjensdeBruyn}.  \\
We can conclude that the majority of the polarized sources are Faraday simple, at least at the wavelengths accessible with the data presented here. The few outliers in Fig. \ref{fig:rmlinearity} follow the $\lambda^2$ behavior when the RM synthesis value is assumed. Therefore, it is not necessary to exclude these sources from our analysis.\\
To summarize, the RM determined with the RM synthesis technique is due to the Faraday screens between the emitting source and the telescope. In principle, there could be several contributions to the RM: the Galaxy, the Fornax cluster ICM, the cosmic web, and the local environment of the emitting source. At these frequencies, the dominant contributions are due to the first two, especially when considering the global properties of the RM, such as the radial profiles. In the next section, we show the observed RM grid and the Galactic subtracted RM grid that should be a close reconstruction of the Fornax cluster RM.

\begin{figure}
    \centering
    \includegraphics[width=0.49\textwidth]{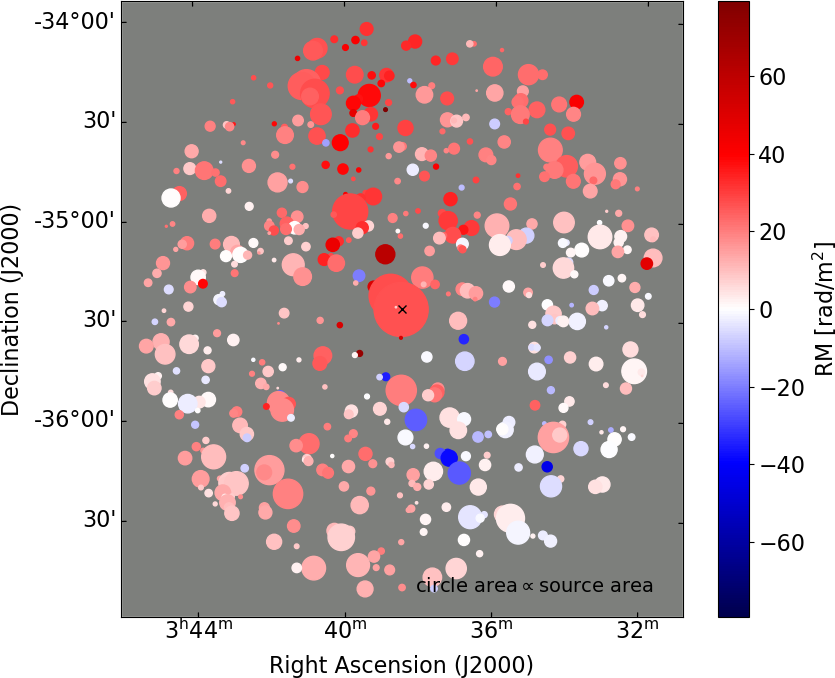}\\
    \vspace{0.3cm}
    \includegraphics[width=0.49\textwidth]{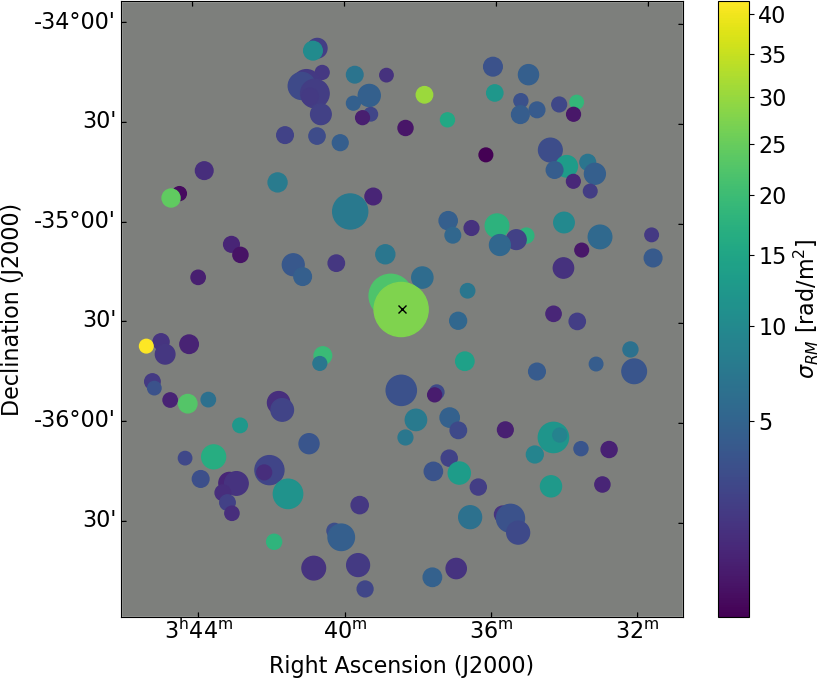}
    \caption{Rotation measure grid evaluated from the MeerKAT Fornax Survey broadband data between 900\,MHz and 1.4\,GHz. The image is centered on the Fornax cluster center and goes out to a radius of 1.42\,deg. Each circle represents a source with the area proportional to the source area. Top: RM mean per source. Bottom: RM standard deviation per source considering sources with cover at least three times the beam area.}
    \label{fig:RMgrid}
\end{figure}

\subsection{RM images}
\begin{figure}
    \centering
    \includegraphics[width=0.49\textwidth]{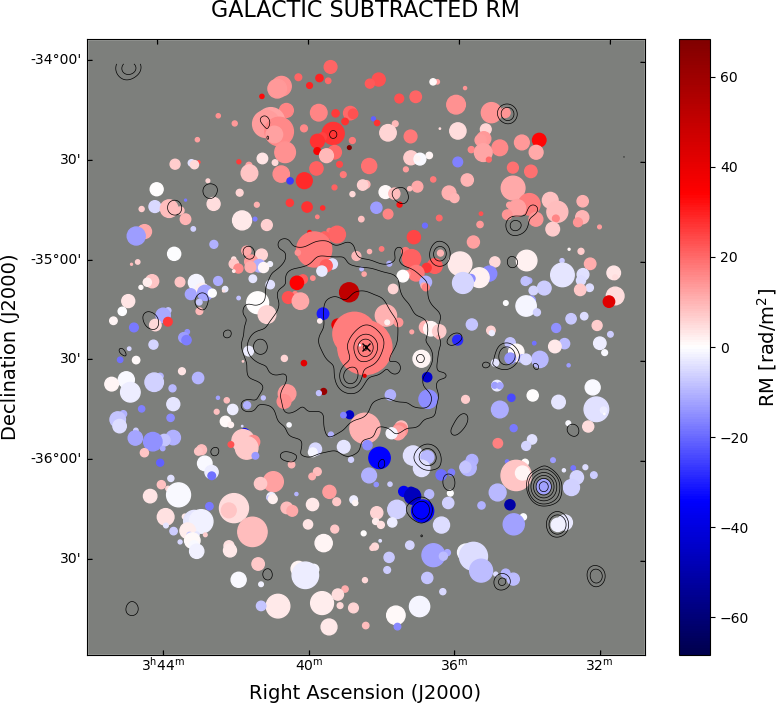}
    \caption{Rotation measure grid after the subtraction of the Galactic RM \cite{anderson2021}. The images are centered on the Fornax cluster center and go down to a radius of 1.42\,deg. Each circle represents a source with the area proportional to the source area, color-coded according to its average RM. The contours refer to the 0.2--2.3\,keV e-ROSITA surface brightness image (smoothed at 5$\arcmin$) and start at 3$\sigma$ with $\sigma$=0.072\,counts s$^{-1}$deg$^{-2}$ and scale with $\sqrt{2}$.}
    \label{fig:gal_sub}
\end{figure}

The RM grid obtained as the Faraday depth at the location of the FDF peak is shown in Fig. \ref{fig:RMgrid} (top). The bottom image shows the RM standard deviation across the source when the source area is larger than three times the beam area. Each circle in the images represents a detected polarized source. The circle area is proportional to the circularized source area\footnote{The image is a scatter plot made with the \texttt{matplotlib} library. We set the size parameter equal to the number of pixels associated with the source. The circle radius is around 18 times the square root of the source area divided by $\pi$.}, while the colors represent the RM mean (top) the standard deviation (bottom) of each source. \\
The RM grid images suggest a RM ordered on large scales with a negative or almost equal to zero trend along SW and positive values going northward. We measure an average RM of 14.7\,rad m$^{-2}$ and a median of 14.3\,rad m$^{-2}$ in the grid. The RM standard deviation image shows relatively low values, indicating an ordered magnetic field. This is also confirmed by the RM standard deviation evaluated in the entire image, which is around 15.3\,rad m$^{-2}$. The above measurements include all the detected sources considering independent resolution elements.\\
In Fig. \ref{fig:gal_sub}, we show the RM grid after the subtraction of the Galactic RM performed with the 2D polynomial reconstruction reported by \cite{anderson2021}. The contours in the figure refer to the 0.2--2.3\,keV eROSITA contours starting at 3$\sigma$, where $\sigma$= 0.072 counts s$^{-1}$ deg$^{-2}$, scaling with a factor of $\sqrt{2}$.
In this figure we can still observe a RM ordered on large scales.
The average and median RM in the image are 4.9 and 4.6\,rad m$^{-2}$, respectively. \\
It is worth noting that \cite{anderson2021} highlight the presence of two distinct morphological subregions on their RM grid. The first is a triangular region of positive and negative RM values with the vertex on NGC\,1399 and its apparent base about $\sim$1.5\,deg to the NE; the second subregion is a negative RM banana-shaped strip of width $\sim$0.5\,deg and length $\sim$2.5\,deg, curving slightly around NGC\,1399, but centered $\sim$0.75\,deg to its SW. With the unprecedented sensitivity and resolution achieved in our data, we can have a more detailed picture of the RM grid.
Indeed, we see a very peculiar and localized positive and negative stripe with high RM values distributed from the cluster center to the N and toward the SSW, respectively. From now on, we shall call this feature the high-RM stripe. In clusters, we expect to see a general decrease in the RM mean value going toward large distances from the cluster center. With the precision of our measurements, we see instead that across the stripe the RM values are higher than what is observed in the field of view regardless of their position with respect to the cluster center. The nature of this intriguing feature will be discussed in the next section.

\begin{figure}
    \centering
    \includegraphics[width=0.4\textwidth]{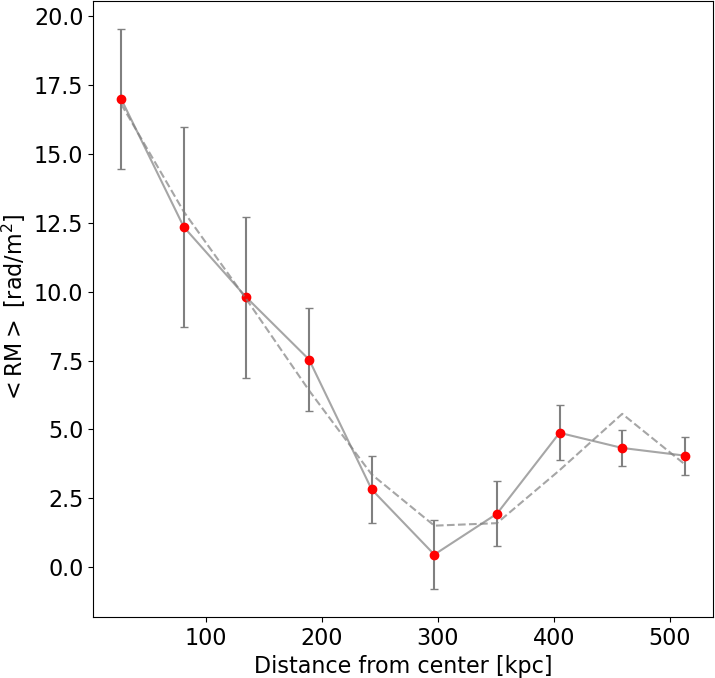}\\
    \vspace{0.3cm}
    \includegraphics[width=0.4\textwidth]{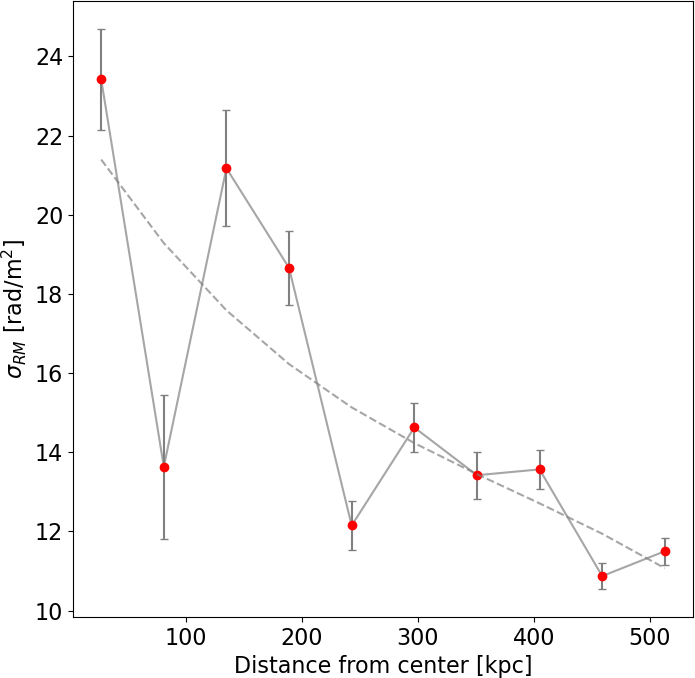}
    \caption{Radial profiles of the RM computed in annuli centered on the cluster center with a width of 8.5$\arcmin$ (51\,kpc). Top: RM mean radial profile. Bottom: RM standard deviation profile.}
    \label{fig:rmprof}
\end{figure}

\begin{figure}
    \centering
    \includegraphics[width=0.4\textwidth]{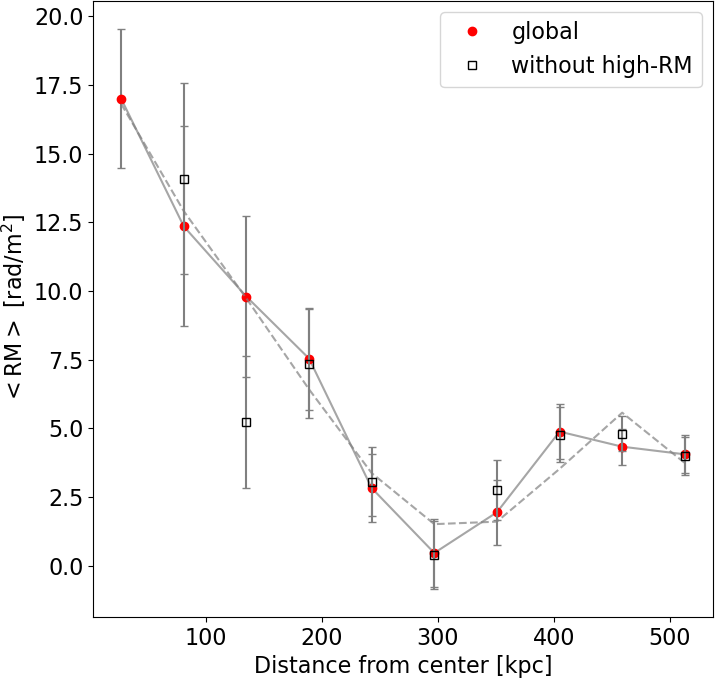}
    \caption{Radial profiles of the RM mean computed in annuli centred on the cluster center with a width of 8.5$\arcmin$ (51\,kpc). Red dots refer to the RM average considering all the sources, while the empty black squares exclude the sources belonging to the high-RM stripe.}
    \label{fig:rmprof_noHRM}
\end{figure}

\subsection{Galactic subtracted rotation measure properties}
We computed the RM radial profiles in annuli centered on the cluster center that have a width of 8.5$\arcmin$; that is, 51\,kpc. The results are shown in Fig. \ref{fig:rmprof}. 
Both the RM mean (top) and the standard deviation (bottom) show the typical RM decrement going from the center to the cluster outskirts, due to a lower intensity of the intra-cluster magnetic field and a less dense thermal plasma.
However, we note two important features.
The RM mean increases at a radial distance of about 300\,kpc from zero to $\sim$5\,rad m$^{-2}$. The RM standard deviation does not go to zero at large distances but rather oscillates around 13\,rad m$^{-2}$. To emphasize these trends, we plot in both the panels the results of a fifth- and third-order polynomial fit in the top and bottom plots, respectively.
As is shown in Fig. \ref{fig:rmprof_noHRM}, the high-RM stripe is not responsible for these features. This plot again shows the RM mean radial profile. The empty black squares represent the profile when the sources belonging to the high-RM stripe are excluded.\\
In their work, \cite{anderson2021} point out the presence of a RM enhancement 2--4 times the projected distance of the X-ray emitting ICM from their RM grid and an excess scatter of about 17\,rad m$^{-2}$ within 1\,deg, which are compatible with the formation of a bow-shock as a consequence of the merging with the Fornax A group on SW. Our data confirm the RM enhancement at two and up to three times the X-ray emitting plasma. Our mosaic does not include more distant regions. We detected a slightly lower RM standard deviation within 1\,deg; that is, $\sim$18\,rad m$^{-2}$, which yields an excess scatter compared to the external region of $\sim$13.8\,rad m$^{-2}$.
The resolution and sensitivity of our data will allow us to further explore all of these features. More details can be found in the next section.\\

%-----------------------------------------------------------------
\section{Discussion}
In Sect. 4, we showed the 2D distribution and radial profile of the RM values and of the RM dispersion. We note a high-RM stripe going from N to SSW across the cluster center, with positive values toward the N and negative values to the S. The RM values across the stripe do not seem to decrease as a function of the distance from the cluster center. We also note an enhancement in the RM mean profile after 300\,kpc and a standard deviation on the order of $\sim$13\,rad m$^{-2}$ at large radii. Here, we want to discuss what is causing these features.\\
Interestingly, there are several aspects that can play a role, since the observed RM is due to the sum of the following contributions:
\begin{enumerate}
    \item the noise,
    \item internal rotation of the polarized sources,
    \item the local environment of the polarized sources,
    \item the Milky Way,
    \item the cosmic web,
    \item the Fornax cluster.
\end{enumerate}

The RM noise is below 2\,rad m$^{-2}$ on average per independent beam and does not show any radial or spatial trend. This means that it is unlikely that a noise fluctuation is causing the RM grid asymmetry. Of course, it is partially contributing to the RM standard deviation plateau.\\

The lack of Faraday complexity excludes a major impact from the internal rotation, at least at this level. Data at longer wavelengths could reveal a complexity hidden at the frequencies that we are considering in this work. However, the quality of the plots shown in the bottom panels of Fig. \ref{fig:rmlinearity} suggests that this scenario is improbable.\\

The local environment is an additional term to the observed RM. Although it is unlikely that a homogeneous distribution of the radio sources can produce an asymmetry in the RM grid and the increase in the RM mean profile, we could expect an impact on the RM standard deviation values: source-to-source variations in the local environment can contribute to the high sigma plateau. In fact, for the most extended sources (sizes larger than ten times the beam area), the average RM standard deviation is around 6\,rad m$^{-2}$. \\

The profiles of Figs. \ref{fig:rmprof} have been computed in the Galactic subtracted RM grid. 
We can consider the possibility of the existence of Galactic RM substructures that were not captured by the modeling proposed by \cite{anderson2021}. However, an inspection of the recent images obtained during the Southern Twenty--centimeter All--sky Polarization Survey \citep[STAPS]{Raycheva2024arXiv240606166R} does not reveal particular features in the Galactic RM image along the high-RM stripe.
An association with Galactic filaments is also questionable due to the Galactic latitude of Fornax (b=$-$54$^{\circ}$). RM coherent structures on top of neutral hydrogen emission have been found on Galactic scales \citep{bracco2020A&A...644L...3B} but an inspection of the Leiden/Argentine/Bonn (LAB) Survey of Galactic HI \citep{Kalberla2005A&A...440..775K} does not indicate the presence of HI galactic filaments in our field. An inspection of the Galactic RM reconstruction by \cite{Hutschenreuter2022A&A...657A..43H} suggests a $\rm\sigma_{RM}\approx6\,rad/m^2$, which can partially explain the $\rm\sigma_{RM}$ plateau at large radii.
To obtain the 13\,rad m$^{-2}$ RM standard deviation observed at large distances from the cluster center, considering the above contributions, we would need an additional term of around 9\,rad m$^{-2}$ as a result of a combination of the cluster and the cosmic web contributions.\\

\begin{figure}
    \includegraphics[width=0.45\textwidth]{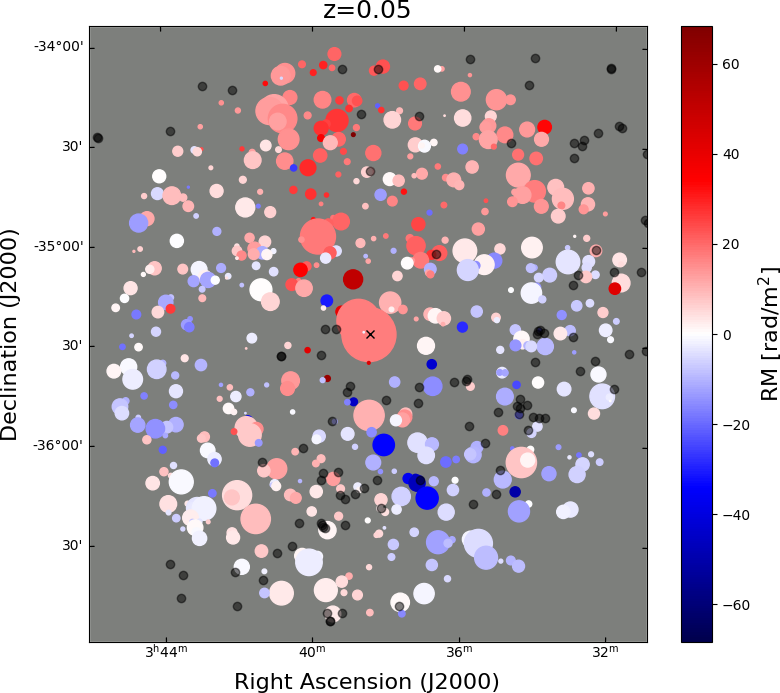}\\
    
    \includegraphics[width=0.45\textwidth]{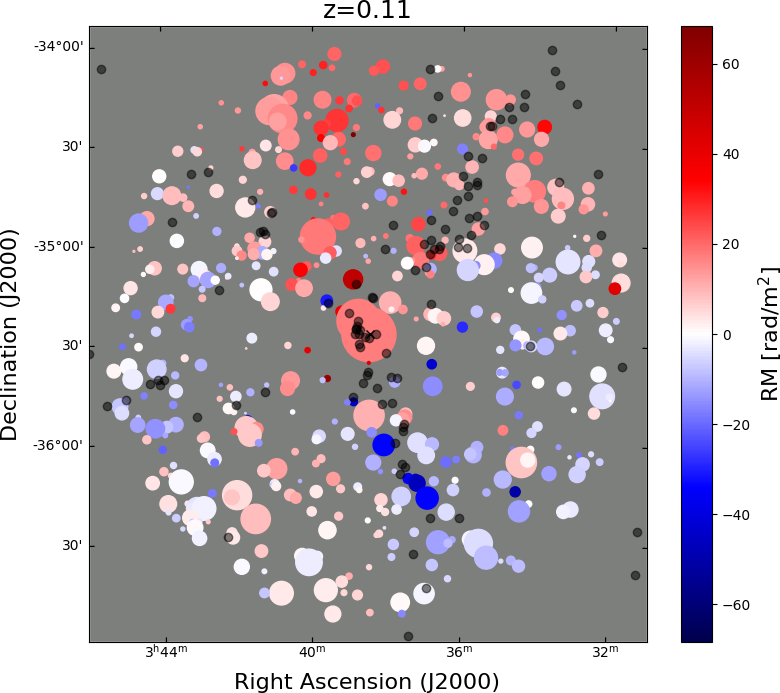}\\
    
    \includegraphics[width=0.45\textwidth]{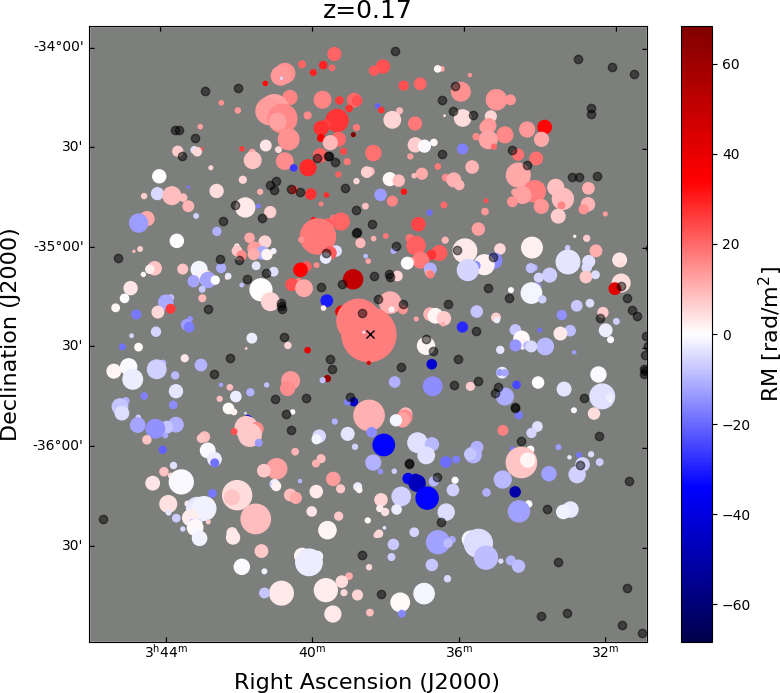}
    \caption{RM grid with the over-densities reported by \cite{maddox} at z=0.05 (top), z=0.11 (middle), and z=0.17 (bottom).}
    \label{fig:RMgrid_fil}
\end{figure}
\begin{figure}
    \centering
    \includegraphics[width=0.45\textwidth]{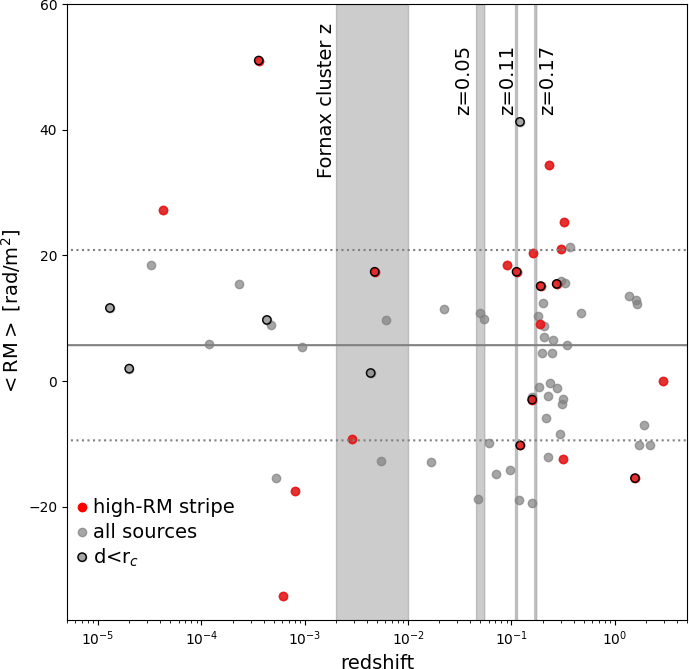}
    \caption{Redshift distribution of the polarized sources.}
    \label{fig:highRMz}
\end{figure}

The RM is indeed contaminated by the cosmic web. Large-scale magnetic fields with a large autocorrelation length can generate high RM contributions even if the magnetic field strength is low. This scenario could explain the high-RM stripe and, to better explore it, we compared our RM catalog with the optical catalog by \citet{maddox}. This work identified in the Fornax field three over-densities with filamentary structure at redshift 0.05, 0.11, and 0.17. We want to understand if the high-RM stripe is caused by these over-densities.\\
Fig. \ref{fig:RMgrid_fil} shows the RM grid image with the sources (black dots) belonging to the z=0.05, z=0.11, and z=0.17 over-densities from top to bottom, respectively.
We do not see a clear spatial coincidence between the high-RM stripe and the z=0.05 over-density. 
The sources at z=0.17 are quite spread out and do not seem to be associated with the high-RM stripe. 
At z=0.11, a filamentary structure is on top of the SW part of the high-RM stripe.
To understand if the high-RM stripe sources are within or in the background of this overdensity, we looked for the redshift identification using the catalog by \cite{maddox}. Among our 508 polarized sources, we found the redshift identification of 71 sources. Fig. \ref{fig:highRMz} shows the mean RM as a function of the source redshift.
In this plot, the sources within the Fornax core radius are plotted with a black contour. We highlight in red the 20 sources belonging to the high-RM stripe. 
The four vertical gray regions show the redshift position and dispersion of the Fornax cluster and of the three over-densities reported by \citet{maddox}. The solid and dotted horizontal gray lines are the median and the standard deviation of the 71 polarized sources with a redshift identification; that is, <RM>=(5.7$\pm$15.1)\,rad m$^{-2}$. Half of the high-RM stripe sources show RM values out of the latter range and are spread in redshift. 
As is shown in this plot, the redshift identification of the RM sources belonging to the high-RM stripe suggests that they are not crossing this filament: it is unlikely that the southern part of the high-RM stripe is due to the z=0.11 over-density.\\
The limited number of redshift identifications (less than 13\% of the RM sample) makes it hard to understand the nature of the high-RM stripe. 
Based on the above considerations, it seems unlikely that the high-RM stripe is associated with one of the three over-densities reported by \citet{maddox}. \\
However, the Southern Sky Redshift Survey \citep{Costa1988ApJ...327..544D} determined that the Fornax cluster is part of a filamentary structure with the Eridanus group (z=0.00557) to the N. The Dorado group is in the same structure to the SE with respect to the Fornax cluster. The two groups are at a distance from the Fornax cluster of 14.88 and 21.69 deg. Moreover, the Fornax A group is located SW from the Fornax cluster center and there could be a merger ongoing between these two structures.\\
It is therefore possible that the high-RM stripe is associated with the large-scale structure (LSS) surrounding the Fornax cluster:
matter grows along the filaments that link the cluster with the surrounding groups and within the virial radius the thermal plasma density is relatively high (higher than that of the over-densities at larger z); this localized excess combined with a quite ordered magnetic field along the filaments and/or an enhancement of the line-of-sight parallel component of the magnetic field caused by the propagation of shock waves could explain the high-RM stripe. Indeed, according to the top-down models of magnetic fields injection, weak ($\sim$nGauss$-$level) magnetic fields are embedded in the LSS of the Universe and merger events are responsible for their amplification and spreading in the dense environment such as the ICM \citep[see e.g.][]{dolag2005MNRAS.358..726D}. The Fornax cluster represents a small-mass and low-density environment; therefore, the dissipation of the large-scale magnetic field into smaller scales could be less efficient.
In a recent paper, \cite{raj2024A&A...690A..92R} reconstruct in detail the LSS surrounding the Fornax cluster. No clear spatial coincident is found between the LSS and the high-RM stripe, probably because they are tracing physical structures that have different scales.
Future works with a larger spatial sampling of the RM are needed to make a fair comparison with the LSS and to confirm the association of the high-RM stripe with the matter accretion in the Fornax cluster. In any case, this seems the most reliable scenario.\\
\begin{figure}
    \centering
    \includegraphics[width=0.45\textwidth]{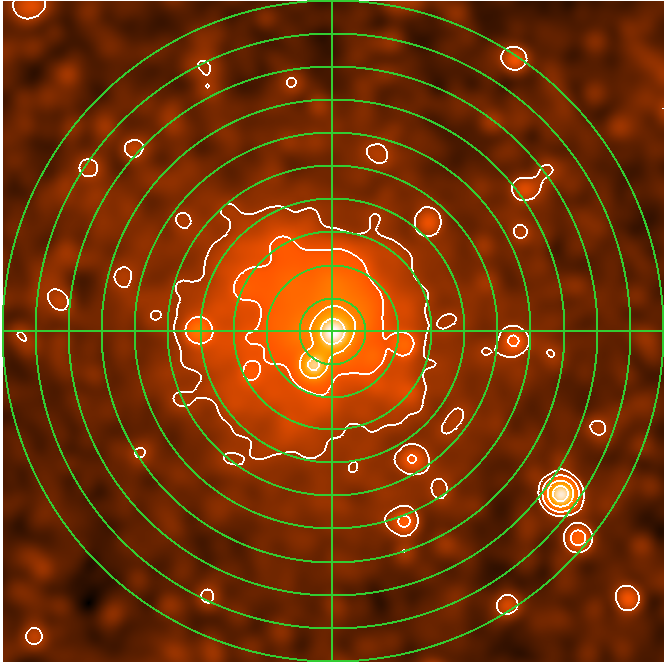}
    \caption{0.2--2.3\,keV SRG/e--ROSITA surface brightness image (smoothed at 5\arcmin) in colors and white contours (same ad Fig.\ref{fig:gal_sub} with the annuli (in green) used to compute the radial profiles.}
    \label{fig:xray_an}
\end{figure}

The RM contributions discussed so far cannot explain the increment of the RM mean radial profile at 300\,kpc.
As we already demonstrated with Fig. \ref{fig:rmprof_noHRM}, the high-RM stripe is not responsible for such behavior. Therefore, it is reasonable that the feature is related to cluster physics. Merger and sloshing events can displace the X-ray emitting gas that contributes to the observed RM. Shock waves propagating as a result of the accretion from surrounding structures can confine the magnetic field and increase, depending on the shock geometry, the strength of its parallel component. The consequence of such phenomena would be an increment of the RM values.\\
The Fornax cluster is experiencing a sloshing event as a consequence of the merging between the NGC\,1399 and NGC\,1404 galaxies \citep{Machacek2005ApJ...621..663M, Scharf2005ApJ...633..154S, Su2017ApJ...851...69S,Sheardown2018ApJ...865..118S}
that is moving the gas toward the east. 
The 0.2-–2.3 keV SRG/e--ROSITA X-ray surface brightness in Fig. \ref{fig:xray_an} indeed shows an X-ray emitting gas that is more confined toward the west. The annuli shown in this plot are the same used to compute the RM radial profiles of Fig. \ref{fig:rmprof}. In this direction, the contours (same as Fig. \ref{fig:gal_sub}) stop at the third anulus, while toward the east the contours reach the fifth anulus.
A merger event is likely occurring with the infalling Fornax A group along the NE--SW axis \citep{Drinkwater2001ApJ...548L.139D} but we cannot see a clear connection between the two structures (the Fornax cluster and the Fornax A group) in the SRG/e--ROSITA image. \\
To better understand how the RM is related with the cluster physics, we show in Figs. \ref{fig:stripe1} and \ref{fig:stripe2} the RM average and standard deviation in different directions.
We considered in each step a box with a width of 18$\arcmin$. This was a compromise to have a significant number of measurements in each box and enough spatial resolution. However, this was not possible in every box, and therefore we plot only those with at least three independent measurements. 
Each profile is color-coded according to the polar angle. In Col. 1, a circle with a color bar is shown. Each row represents a crossing direction, which is indicated in Col. 1 on top of the circle with the color bar.
The RM mean and standard deviation are plotted in Cols. 2-3, respectively. We include in each plot three profiles: one along the crossing direction indicated in Col. 1 and two additional directions separated from it by $\pm$10$^\circ$.\\
\begin{figure*}
        \centering
    \includegraphics[width=0.4\textwidth]{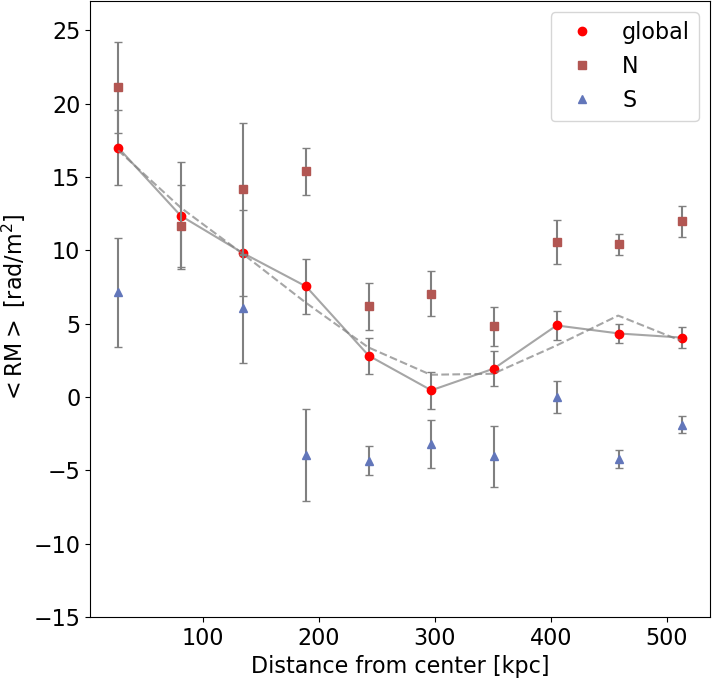}
    \hspace{0.3cm}
    \includegraphics[width=0.4\textwidth]{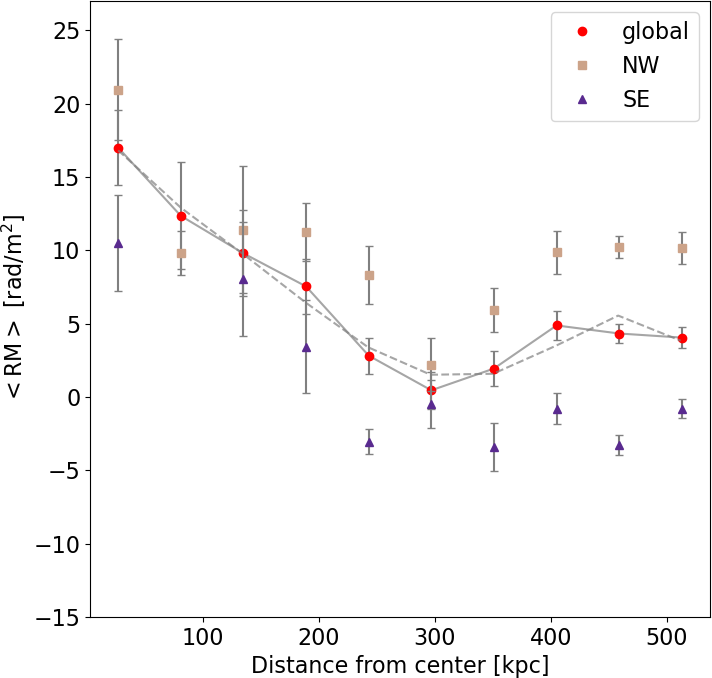}\\
    \vspace{0.3cm}
    \includegraphics[width=0.4\textwidth]{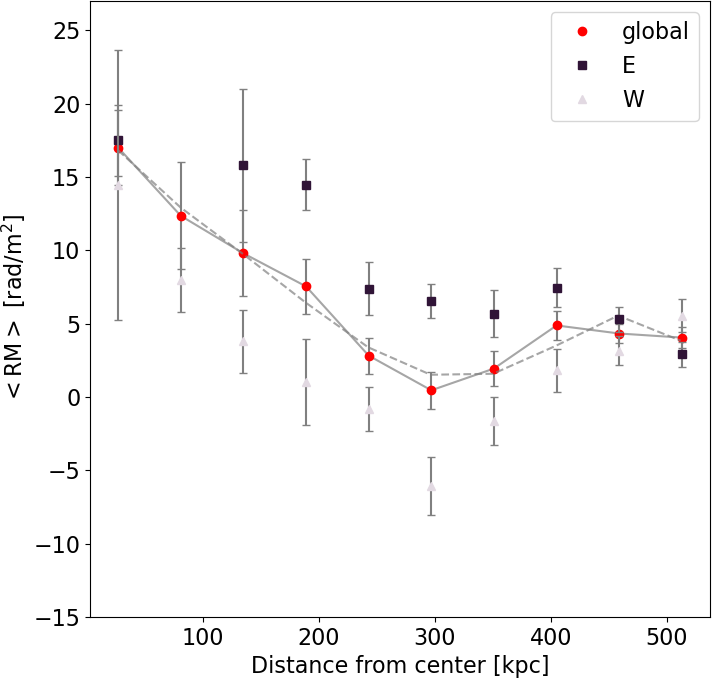}
    \hspace{0.3cm}
    \includegraphics[width=0.4\textwidth]{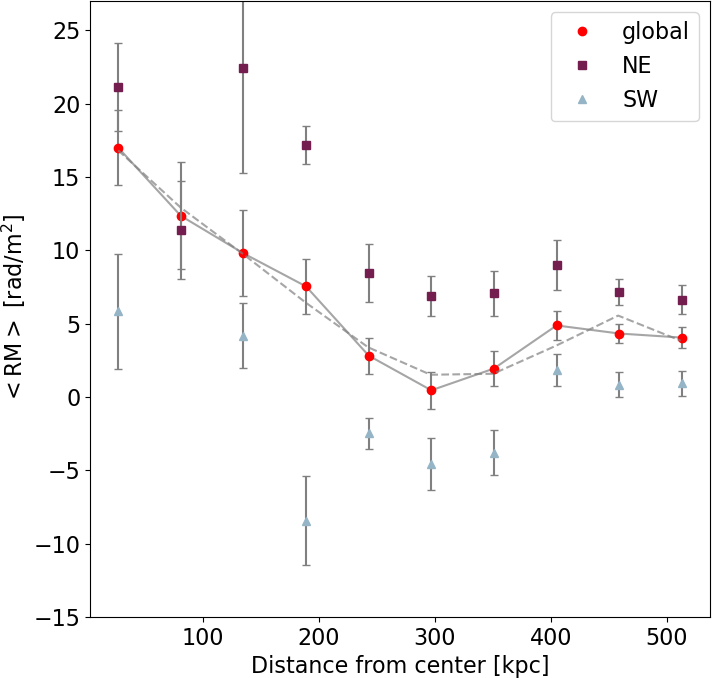}
    \caption{ Radial profiles of the RM computed in semi-circular annuli with a width of 8.5$\arcmin$ (i.e. 51\,kpc). The points are color-coded according to the direction of the section indicated in the top right corner of each panel. Top left: RM mean radial profile of the N and S sectors. Top right: RM mean radial profile of the NW and SE sectors. Bottom left: RM mean radial profile of the east and west sectors. Bottom right: RM mean radial profile of the NE and SW sectors.}
    \label{fig:rmprof_dir}
\end{figure*}
The RM standard deviation profiles do not show any clear hint of asymmetry regardless of the crossing direction.
On the other hand, it is interesting to note a clear asymmetry in the RM radial profiles that reaches its maximum crossing the cluster from NNE to SSW. In this direction, we can see positive RM values that decrease to negative values once we cross the cluster center. This is due to the presence of the high--RM stripe. Another and even more interesting trend is along the SE--NW and SSE--NWW directions, where the RM average goes from negative to positive values. The sloshing event is affecting the cluster toward the east, the merging event could be undergoing along the SW and the accretion is likely happening toward the N and SW. Nothing is happening toward the NW and W. The high RM values in this direction could be due to a larger autocorrelation length of the Fornax cluster magnetic field: in the opposite direction, the small-scale turbulence is causing a decay of the magnetic field power spectrum toward smaller and smaller scales. Such behavior is observed in numerical simulations \citep[see e.g. Fig. 14 in][]{vacca2024A&A...691A.334V}. As a consequence, the RM values are lower toward the east, while they do not show a decrease toward the west. \\
Fig. \ref{fig:rmprof_dir} shows the RM mean profiles computed across N and S (top left), NW and SE (top right), E and W (bottom left), and NE and SE (bottom panels) sectors. In each panel, the global profile of Fig. \ref{fig:rmprof} is shown for comparison (red dots).
It is clear that the RM mean increases at large radii, especially across the N and NW sectors. Since the same increment is not detected in the NE sector, we can conclude that the feature observed in the global profile is mainly due to the RM properties toward the west. As was already pointed out, in this part of the cluster there are no merger, sloshing, or accretion phenomena ongoing. Our interpretation of the increment of the RM mean radial profile starting at 300\,kpc is therefore a magnetic field in the western part of the cluster with a structure closer to the injected field and not an evolved magnetic field. The Fornax cluster magnetic field in this region could be characterized by an autocorrelation length that is higher than what is observed in the opposite part of the cluster, where turbulence is dissipating the field into smaller scales.\\

%-----------------------------------------------------------------
\section{Summary and conclusions}
In this work, we report the first results of the MeerKAT Fornax Survey broadband polarized observations.
This is the deepest survey ever to have been carried out in polarization at mid-frequencies.
The methods applied to derive the polarization products allowed us to reconstruct the densest RM grid ever built, $\sim$80 polarized sources per deg$^2$, comparable with expectations for the polarization survey planned with SKA1-MID \citep{Heald2020Galax...8...53H}. 
Of the 508 polarized sources detected, five are cluster sources. The differential polarized source counts are in good agreement with previous surveys and are probing the faint radio sky even at lower polarized flux densities; that is, a factor of ten deeper with respect to previous surveys. At low flux density, we observe for the first time an increase in the differential source counts that is compatible with AGN populations rather than star-forming galaxies. \\
The properties of the RM are in good agreement with what is reported by \cite{anderson2021}. We detect the decrement of the RM average and standard deviation as a function of distance from the cluster center, the increment of the RM scatter beyond 300\,kpc from the cluster center and standard deviation plateau at large radii. We also detect for the first time a high-RM stripe in the RM grid from N to SSW. Thanks to the density of the RM grid presented in this work, we could investigate the RM properties with an unprecedented spatial resolution. 
The main result of this work is that the RM is a powerful tool to understand the Fornax galaxy cluster physics:
\begin{enumerate}
    \item first, it is tracing different properties of the cluster magnetic field, depending on the phenomena ongoing locally. In the eastern part of the cluster, sloshing phenomena are driving the evolution of the cluster magnetic field, whose power spectrum is decaying into smaller and smaller scales, resulting in lower values of the RM. In the opposite part, the lack of dynamical motion is helping the cluster magnetic field to retain an autocorrelation length larger with respect to the other regions, resulting in a larger RM. As a result, we see an increment in the RM mean radial profile due to the combination of these effects;
    \item second, the high-RM stripe is likely due to the accretion of matter into the Fornax galaxy cluster along the surrounding cosmic web filaments. Shock waves propagating in these directions can confine the magnetic field, enhancing its strength and compressing the matter. As a result, we detect RM values that are higher than expected.
\end{enumerate}
In the next paper, starting from these conclusions, we shall perform a measurement of the Fornax cluster magnetic field. 

\begin{acknowledgements}
We greatly thank the anonymous referee and the editor for their constructive comments, which have greatly enhanced the quality of the paper. We thank Dr. Ang Liu and Dr. Rosita Paladino for their help.
We are grateful to the full MeerKAT team for their work building, commissioning and operating MeerKAT, and for their support to the MeerKAT Fornax Survey.
The MeerKAT telescope is operated by the South African Radio Astronomy Observatory, which is a facility of the National Research Foundation, an agency of the Department of Science and Innovation.
This project has received funding from the European Research Council (ERC) under the European Union’s Horizon 2020 research and innovation programme (grant agreement no. 679627, “FORNAX”; and grant agreement no. 882793, “MeerGas”). 
The data of the MeerKAT Fornax Survey are reduced using the CARACal pipeline, partially supported by ERC Starting grant number 679627, MAECI Grant Number ZA18GR02, DST-NRF Grant Number 113121 as part of the ISARP Joint Research Scheme, and BMBF project 05A17PC2 for D-MeerKAT. Information about CARACal can be obtained online under the URL: https://caracal.readthedocs.io. 
FL acknowledges financial support from the Italian Ministry of University and Research – Project Proposal CIR01-00010.
VV acknowledges support from the Premio per Giovani Ricercatori "Gianni Tofani" II edizione, promoted by INAF-Osservatorio Astrofisico di Arcetri (DD n. 84/2023).
PK is partially supported by the BMBF project 05A23PC1 for D-MeerKAT III.
This work is based on data from eROSITA, the soft X-ray instrument aboard SRG, a joint Russian-German science mission supported by the Russian Space Agency (Roskosmos), in the interests of the Russian Academy of Sciences represented by its Space Research Institute (IKI), and the Deutsches Zentrum für Luft- und Raumfahrt (DLR). The SRG spacecraft was built by Lavochkin Association (NPOL) and its subcontractors, and is operated by NPOL with support from the Max Planck Institute for Extraterrestrial Physics (MPE). The development and construction of the eROSITA X-ray instrument was led by MPE, with contributions from the Dr. Karl Remeis Observatory Bamberg \& ECAP (FAU Erlangen-Nuernberg), the University of Hamburg Observatory, the Leibniz Institute for Astrophysics Potsdam (AIP), and the Institute for Astronomy and Astrophysics of the University of Tübingen, with the support of DLR and the Max Planck Society. The Argelander Institute for Astronomy of the University of Bonn and the Ludwig Maximilians Universität Munich also participated in the science preparation for eROSITA.
The eROSITA data shown here were processed using the eSASS software system developed by the German eROSITA consortium.
\end{acknowledgements}

\bibliographystyle{aa}
\bibliography{fornax}

\begin{appendix}
\section{Polarization calibration with CARACal}
\label{app:polcal}
The \texttt{CARACal} pipeline includes three strategies to calibrate the polarization, depending on the calibrators available during the observations. 
The first method (PCAL 1) consists in calibrating the on-axis leakage with an unpolarized source and the cross-hand phase and delay with a known polarized source (i.e. 3C138 or 3C286). The second one (PCAL 2) calibrates all the previous terms with a known polarized source (i.e. 3C138 or 3C286). The third one (PCAL 3) calibrates with a polarized source observed at different parallactic angle (PA) without assuming a model. The model is indeed derived by fitting the QU signals against the PA. This method is particularly useful because there are only two well known polarized sources in the southern sky at limited LSTs (even if SARAO is monitoring new polarized sources). \\
We test all of these strategies on a commissioning 4k data set (ID: 1600463770$\_$sdp$\_$l0, P.I.: Paolo Serra, taken on 18-Sep-2020) which contains seven scans of 3C138. After the calibration of the parallel feeds we derive the polarization calibration correction in the three different ways described above, using the first scan(s) for the PCAL 1(2) method and assuming the model reported in the NRAO webpage.
The three sets of solutions are then applied to the last scans of 3C138 which are then imaged in 68 frequency channels of 10\,MHz each. 
\begin{figure}[h!]
    \centering
    \includegraphics[width=0.47\textwidth]{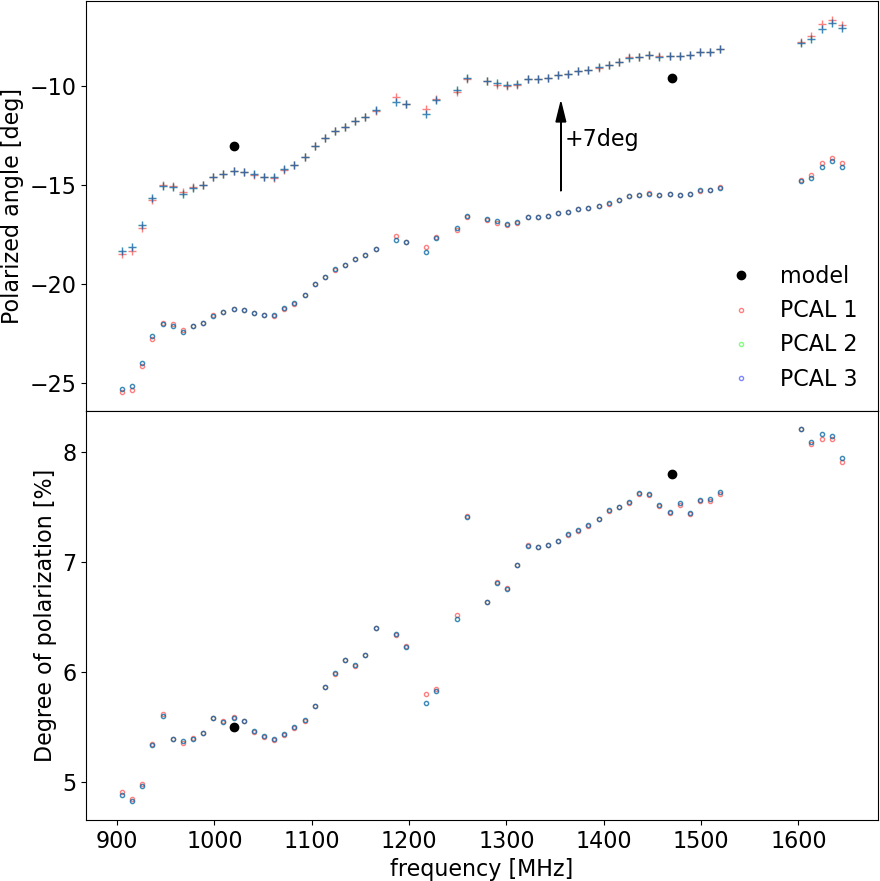}
    \caption{Measured polarization properties of 3C138. Top and bottom panels show the polarized angle and the degree of polarization, respectively. We show the model (black dots), MeerKAT data (dots) color-coded depending on the calibration strategy. In the top panel we also show the MeerKAT data plus 7\,deg (crosses).}
    \label{fig:polcal}
\end{figure}
We show the polarized angle and the degree of polarization measured by fitting the I, Q, and U Stokes parameter in the top and bottom panels of Fig. \ref{fig:polcal}, respectively. Black dots show the model while the data are color-coded depending on the calibration strategy. The solutions are stable regardless of the method we applied.
We also show the measured polarized angles with the addition of a 7\,deg offset. 
Offsets in the polarized angles of MeerKAT images have been reported by many users. According to our analysis they are not compatible with ionospheric effects. Using the \texttt{RMextractor} tool \citep{Mevius2018ascl.soft06024M} we found corrections of $\sim$0.3\,rad m$^{-2}$, i.e. less than 1\,deg. \\
During our observation campaign we establish that the offsets are quite stable in time and therefore they do not represent an issue to our analysis since we are interested in relative measurements rather than absolute quantities.

\section{Is there a deficit of polarized sources?}
In Fig. \ref{fig:pol} the sources do not seem to be homogeneously distributed, else it seems that there is a deficit of sources especially in the central part eastward. This could be due to dynamic range limitation or depolarization effect. We investigated if the lack of sources is real and what is causing it by evaluating the density of polarized sources as a function of the radial distance from the cluster center. 
\begin{figure}
    \centering
    \includegraphics[width=0.47\textwidth]{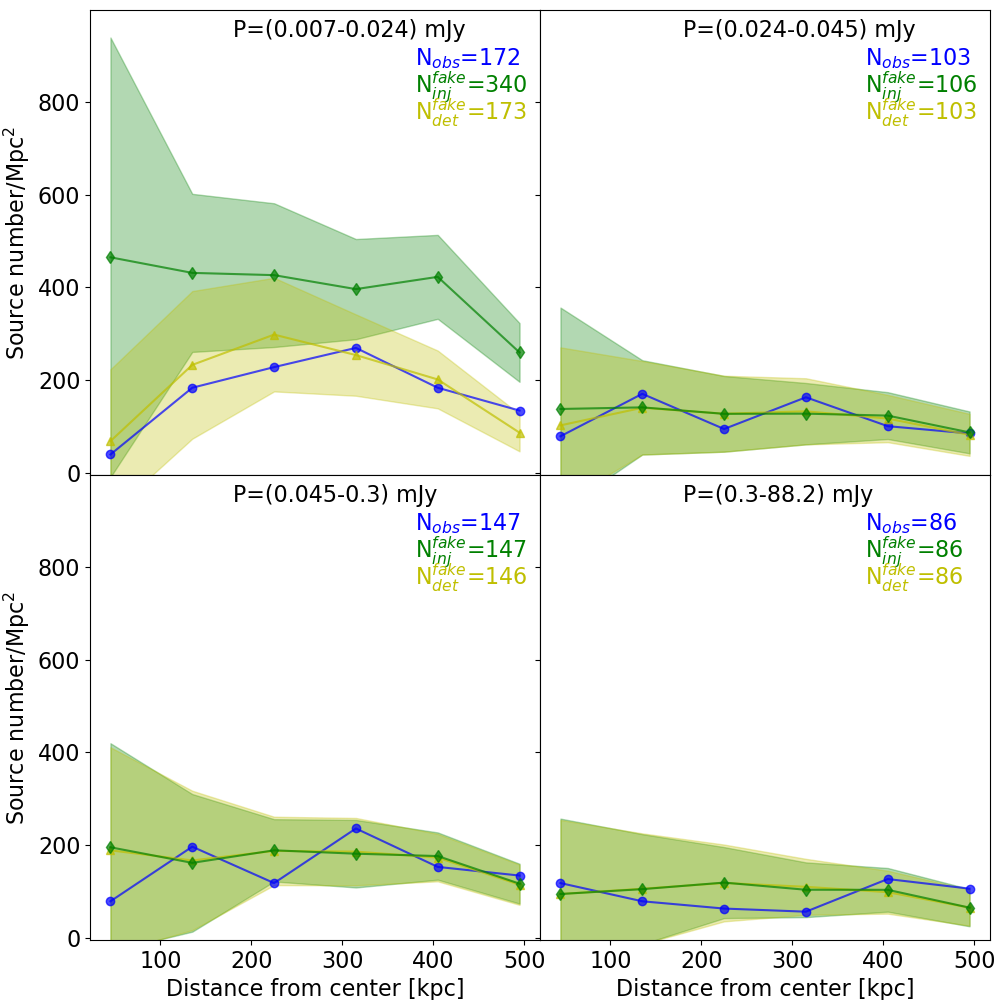}
    \caption{Number of sources per annuli area as a function of the radial distance from the cluster center. Each panel shows a different polarized flux density interval indicated on the top. The data, fake injected and detected source density is shown in blue, green, and yellow, respectively.}
    \label{fig:dens_r}
\end{figure}
\FloatBarrier
We split the sources considering four polarized flux density intervals reported on the top left corner of Fig. \ref{fig:dens_r}. Blue points and solid lines show the data. For each interval we ran a simulation. As done in Sect. 3.3 to evaluate the completeness, we injected fake sources in the Q and U de-rotated Stokes images and we repeated the source detection loop to detect the fake sources. For each polarized flux density interval we repeated the injection and the detection 30 times. The number of injected sources is established in order to obtain a fake detection number equal to the real detections at every polarized flux density range. All of these numbers are shown in the top right corner of each panel.
We plot the injected and detected average source density among the 30 simulated runs as green and yellow dots and solid lines, respectively. A shaded area shows the 3$\sigma$ threshold centered at the average source density, where sigma is the standard deviation computed within each annulus among the 30 simulations. 
In every panel the injected source density is flat by construction. Between 0.024 and 0.3 mJy of polarized flux density we see a fluctuation of the data. Such fluctuation does not exceed the 3$\sigma$ limit of the fake source density distribution which appears quite flat. Between 0.007 and 0.024\,mJy the simulated distribution is not flat and the data follows this behavior. Such a trend is likely due to the noise increase toward the field center because of dynamic range limitation. We remind the reader that we are not simulating the bandwidth depolarization. 
At large polarized flux density (bottom right panel) the data shows a decrement, between 150 and 300 kpc, and an increment, at around 400 kpc, with respect to the average of simulated source density distributions. Nevertheless, they do not exceed the 3$\sigma$ limit of the fake source density distribution. \\
To conclude, the source density in polarization shows a deficit at low flux density (between 0.007 and 0.024 mJy) toward the center which is likely due to the noise distribution. Above 0.024 mJy the deficit is not significant.

\section{Additional figures}
\begin{figure}[h!]
\centering\includegraphics[width=0.7\textwidth]{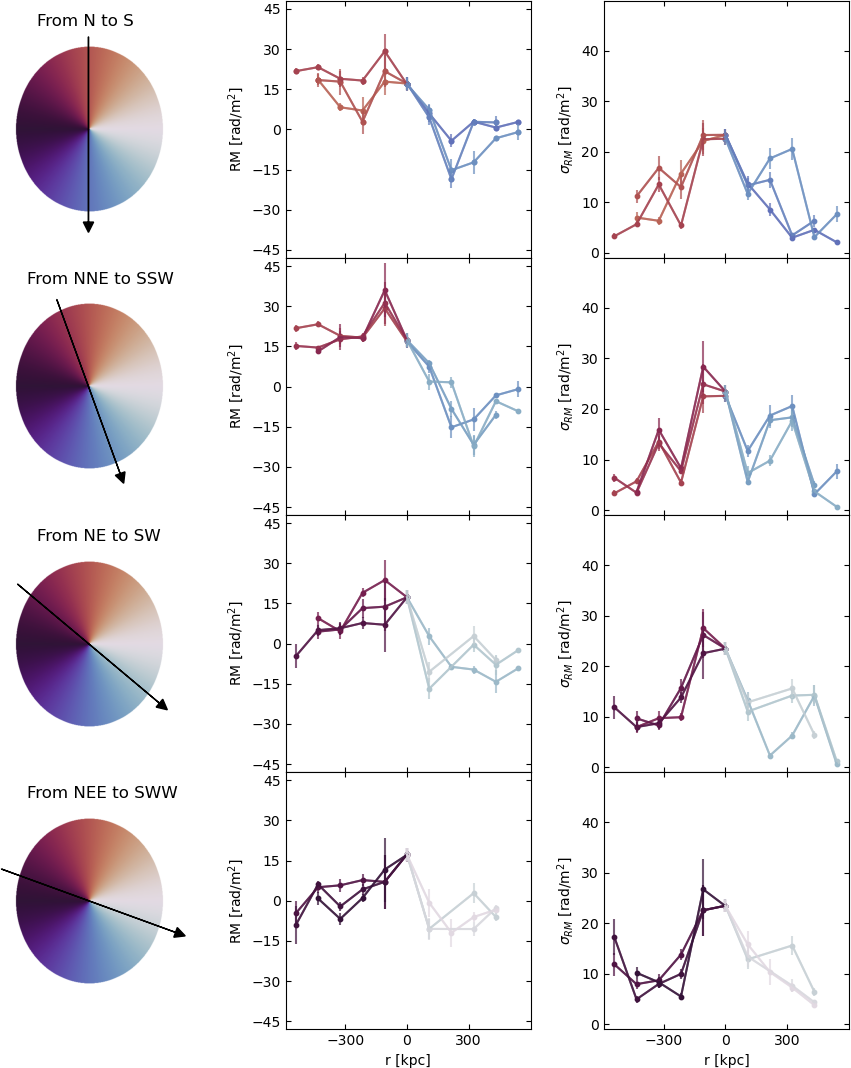}
\caption{RM trend across several directions. Column 1 shows the color bar for crossing direction (black arrow). Columns 2-3 show the mean and standard deviation RM across the corresponding direction and $\pm$10$^\circ$, computed in 18$\arcmin$ square boxes.}
\label{fig:stripe1}
\end{figure}
\FloatBarrier

\begin{figure*}
\includegraphics[width=0.7\textwidth]{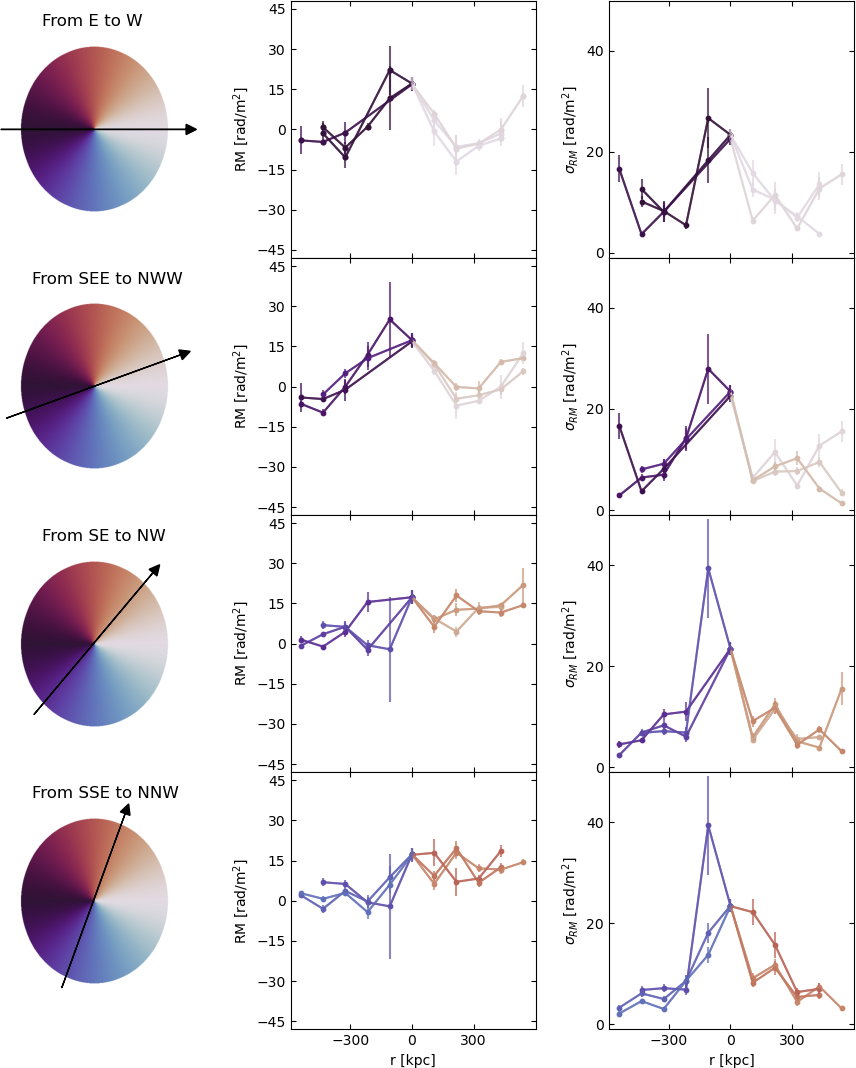}
\caption{RM trend across several directions. Column 1 shows the color bar for crossing direction (black arrow). Columns 2-3 show the mean and standard deviation RM across the corresponding direction and $\pm$10$^\circ$, computed in 18$\arcmin$ square boxes.}
\label{fig:stripe2}
\end{figure*}

\end{appendix}
\end{document}